\documentclass[letterpaper]{emulateapj}
\usepackage{apjfonts}
\usepackage{lscape}
\usepackage{graphicx}
\shorttitle{Black Holes in Dense Star Clusters}
\shortauthors{O'Leary et al.}

\newcommand{\Fewbody}{{\tt Fewbody}}

\begin{document}

\title{Binary Mergers and Growth of Black Holes in Dense Star Clusters}

\author{Ryan M.\ O'Leary}
\author{Frederic A.\ Rasio}
\author{John M.\ Fregeau}
\author{Natalia Ivanova}
\author{Richard O'Shaughnessy}
\affil{Northwestern University, Department of Physics and Astronomy, 2132 Tech Drive, Evanston, IL 60208;\\
oleary@alumni.northwestern.edu, rasio@northwestern.edu, fregeau@northwestern.edu, nata@northwestern.edu, oshaughn@northwestern.edu}

\begin{abstract}
We model the dynamical evolution of primordial black holes (BHs) in
dense star clusters using a simplified treatment of stellar dynamics
in which the BHs are assumed to remain concentrated in an inner core,
completely decoupled from the background stars.  Dynamical
interactions involving BH binaries are computed exactly and are
generated according to a Monte Carlo prescription. Recoil and
ejections lead to complete evaporation of the BH core on a timescale
$\sim 10^9\,$yr for typical globular cluster parameters.  Orbital
decay driven by gravitational radiation can make binaries merge and,
in some cases, successive mergers can lead to significant BH growth.
Our highly simplified treatment of the cluster dynamics allows us to
study a large number of models and to compute statistical
distributions of outcomes, such as the probability of massive BH
growth and retention in a cluster.  We find that, in most models,
there is a significant probability ($\sim 20 - 80\%$) of BH growth
with final masses $\ga 100\,M_{\odot}$. In one case, a BH formed with
mass $\approx 620\,M_{\odot}$.  However, if the typical merger recoil
speed (due to asymmetric emission of gravitational radiation)
significantly exceeds the cluster escape speed, no growth ever
occurs. Independent of the recoil speed, we find that BH-BH mergers
enhanced by dynamical interactions in cluster cores present an
important source of gravitational waves for ground--based laser
interferometers.  Under optimistic conditions, the total rate of
detections by Advanced LIGO, could be as high as a few tens of events
per year from inspiraling BHs from clusters.
\end{abstract}


\section{Introduction}
\label{intro}
\subsection{Astrophysical Motivation}
\label{astrointro}
Many observations of globular clusters suggest the possible
existence of intermediate-mass black holes (IMBHs) with masses 
$\sim10^3 - 10^4\,M_\odot$ in the centers of some cluster
cores. The predicted masses of these IMBHs agree well with a simple extrapolation
of the $M-\sigma$ (mass -- velocity dispersion) relation for galactic nuclei
\citep{2000ApJ...539L..13G}. Observations and dynamical modeling of the globular
clusters M15 in the Milky Way and G1 in M31 appear to be consistent
with such a central massive BH (\citealt{2002AJ....124.3270G,2003AJ....125..376G}; \citealt*{2002ApJ...578L..41G}).
However, $N$-body
simulations by \citet{2003ApJ...582L..21B,2003ApJ...589L..25B} suggest that the
observations of M15 and G1, and, in general, the properties of all
{\it core-collapsed \/} clusters, could be explained equally well by
the presence of many compact stars near the center without a massive
BH (cf.\ \citealt*{2004cbhg.symp...37V,grh05}).
On the other hand, these $N$-body
simulations also suggest that many, perhaps most, {\it non-core-collapsed \/} clusters
(representing about 80\% of globular clusters in the Milky Way) could
contain a central IMBH \citep{2004ApJ...613.1143B,2005ApJ...620..238B}.

Although the origin of
IMBHs is not directly constrained by any observations, one possibility
that has received considerable attention is the growth of a very massive
object through successive collisions and mergers of ordinary massive stars
in the cluster core \citep{1967ApJ...150..163C, 2001ApJ...562L..19E}. Recent numerical studies by
\citet{2002ApJ...576..899P}, \citet*{2004ApJ...604..632G}, and \citet*{2005astro.ph..3130F} demonstrate that mass
segregation and core collapse could proceed so quickly that there is
rapid, runaway growth through stellar collisions before the most
massive stars have evolved to supernovae (in $\la 3\,$Myr). Other
numerical evidence comes from the $N$-body simulations of
\citet{2004Natur.428..724P}, which demonstrate such growth in some young clusters.

An important alternative, which we study in this 
paper, is the growth of an IMBH through successive mergers of stellar mass
($\sim10\,M_\odot$) BHs \citep{1989ApJ...343..725Q,1995MNRAS.272..605L,2001CQGra..18.3977L}.  In this case, the massive stars formed initially
in the cluster (with masses $\ga 20\,M_\odot$) must have avoided physical 
collisions and mergers and instead were able to complete their normal
stellar evolution and produce BHs. Unlike massive stars, BHs
have negligible cross sections for direct collision, so BH mergers can only
occur through gravitational wave (GW) emission in binaries, possibly enhanced 
by dynamical interactions in the cluster (see \citealt{2004IJMPD..13....1M} for a
review and \S \ref{bhform}).  

Even if it does not lead to growth and IMBH formation, the dynamical evolution of
BHs in dense star clusters could also produce large numbers of tight BH--BH
binaries that will merge through GW emission (possibly outside the cluster following
dynamical ejection). These merging BH--BH binaries are 
likely candidates for direct GW detection and could even dominate the 
detection rates for ground--based laser interferometers such as LIGO \citep{2000ApJ...528L..17P}.

\subsection{BH Formation and Segregation}
\label{bhform}

In a globular cluster, it can be expected that a fraction $\sim
10^{-6}$ to $10^{-4}$ of the $N \sim 10^{6}$ initial stars will become
stellar-mass BHs via normal stellar evolution
\citep{1993Natur.364..423S}. Assuming that all stars with initial mass greater than
$20\,M_{\odot}$ become BHs, the most recent Kroupa initial mass
function (IMF) \citep{2003ApJ...598.1076K} gives a slightly higher initial BH
fraction of $N_{\rm BH} \approx 1.5\times 10^{-3} N$, where we use
$m_{\rm min} = .08\,M_{\odot}$ and $m_{\rm max} = 150\,M_{\odot}$ as 
the minimum and maximum stellar masses.  When we scale
this to the total mass of the cluster, $M_{\rm cl}$, we find $N_{\rm
BH} \approx 3\times 10^{-3} (M_{\rm cl}/M_{\odot})$. All of these BHs
should have formed before $\sim 10\,$Myr, with the most massive BHs
forming at around $3\,$Myr \citep{1992A&AS...96..269S}.

 The radial distribution of these BHs in the cluster when they form is
not known, but it is reasonable to assume that they should be much more
centrally concentrated than the remaining main-sequence stars (MSs).
This is for three reasons: (1) we expect significant mass
segregation of the initial, higher-mass progenitors \citep{2005astro.ph..3130F}; (2)
there may be preferential formation of massive stars near the 
cluster center \citep[see,
e.g.,][]{1996ApJ...467..728M,2001MNRAS.323..785B}; (3) BH birth kicks are not expected to be large
enough to displace the BHs into the cluster halo (or eject them from the 
cluster; see \citealt{1996ApJ...473L..25W,2005ApJ...625..324W}).
Even if the BHs were initially distributed throughout the cluster,
mass segregation would still be able to concentrate them into a central 
sub-cluster on a relatively short timescale. Indeed, a BH of 
mass $M_{\rm BH}$ near the half-mass radius will be brought into the 
cluster core on the timescale
\begin{equation}
t_{\rm{seg}} \sim \frac{\left<m\right>}{M_{\rm BH}} t_{\rm{rh}},
\label{eqseg}
\end{equation}
 where $t_{\rm{rh}}$ is the relaxation time at the half-mass radius,
 and $\left<m\right>$ is the average stellar mass \citep{2002ApJ...570..171F}.
 Considering a typical cluster with $t_{\rm rh}~\sim 1\,$Gyr, we
 see that a sub-cluster of BHs should still assemble near the center after
 at most $\sim 100 \,$Myr.
  
After a time very short compared to the overall dynamical evolution timescale
of the globular cluster, the BHs should then form a self-gravitating
subsystem within the core, which is dynamically decoupled from the
rest of the cluster. This decoupling is sometimes referred to as Spitzer's
``mass stratification instability'' \citep{1969ApJ...158L.139S}. The physics of this instability
is by now very well understood, both for simple two-component systems
and for clusters with a broad mass function \citep{2000ApJ...539..331W,2004ApJ...604..632G}.

The dynamical evolution of the BH subsystem proceeds on a much shorter timescale
since its relaxation time is now $\sim~N_{\rm BH}/N$ times smaller than for
the parent cluster. Interactions involving hardening of BH binaries 
will eventually lead to the ejection of
BHs from the cluster, until there are so few left that the BH subsystem
recouples dynamically (returns to ``thermal equilibrium'') with the other 
cluster stars, and the interaction rate 
between BHs and cluster stars becomes comparable to the BH--BH
interaction rate. For simple two-component clusters, \citet{1969ApJ...158L.139S} derived
through analytic methods the condition necessary to reach
energy equipartition:
\begin{equation}
\left(\frac{M_2}{M_1}\right)\left(\frac{m_2}{m_1}\right)^{1.5} < 0.16,
\label{eqpartspit}
\end{equation}
 where $M_2 < M_1$ are the total masses of the two components, and
 $m_2 > m_1$ the mass of each individual object. \citet*{2000ApJ...539..331W} used
 an empirical approach to find the more accurate condition
\begin{equation}
 \left(\frac{M_2}{M_1}\right)\left(\frac{m_2}{m_1}\right)^{2.4} < 0.32.
\label{eqpart}
\end{equation}
For a cluster with total mass $M_1 = 10^6\,M_{\odot}$, BH mass $ m_2 =
15\,M_{\odot}$, and average star mass $m_1 = 1\,M_{\odot}$, the
cluster can be in equipartition, according to equation~(\ref{eqpart}),
when there are $\la 30$ BHs in the cluster, i.e., significantly fewer
than the number expected to form from the IMF. The minimum number of
BHs required to decouple from the cluster would likely be even less if
their mass spectrum was considered \citep{2004ApJ...604..632G}.

For most of its subsequent dynamical evolution, we expect the BH
sub-cluster to remain largely free of other kinds of stars, even MSs
with comparable masses. Indeed, for BHs to have formed, the most
massive stars in the cluster must have avoided runaway collisions.
This requires an initial half-mass relaxation time $t_{\rm rh}\ga
30\,$Myr \citep{2004ApJ...604..632G}.  On the other hand, by the time
the MS turnoff has decreased to $10\,M_{\odot}$, driving core collapse
to concentrate the remaining $10\,M_{\odot}$ MSs into the core (where
they could then interact with the BHs) before they evolve would
require $t_{\rm rh}\la 20 \,$Myr (\citealt{2004ApJ...604..632G}; see
especially their Fig.~10).  Therefore we see that the parameter space
for clusters where both BHs and massive MSs would segregate and
decouple simultaneously is probably very small, or nonexistent. This
will allow us to concentrate on ``pure'' BH systems in our numerical
simulations (\S 2). This picture is not altered significantly by the
presence of binaries.  During mass segregation, as the BHs start
concentrating into the denser cluster core, they will likely interact
with each other to form BH--BH binaries, if they were not already in
binaries.  BH--MS binaries will quickly undergo three-body or
four-body exchange interactions that replace the lighter MS companion
by a heavier BH \citep{1993ApJ...415..631S}.  Therefore, as the BH
sub-cluster begins to dynamically decouple from the rest of the
cluster, a considerable number of hard BH--BH binaries are expected to
remain.  This has been demonstrated with direct $N$-body simulations
by \citet{2000ApJ...528L..17P}, where they found that almost all
BH-binaries ejected were BH--BH.

\subsection{Previous Studies}

Motivated by the absence of BH X-ray binaries,
\citet*{1993Natur.364..421K} and \citet{1993Natur.364..423S} discussed
the evolution and fate of $10\,M_{\odot}$ BHs in globular clusters
using simple analytic considerations.  Their conclusions can be
summarized as follows.  Through dynamical interactions, hard BH--BH binaries tend to be hardened further
(whereas soft binaries tend to be disrupted;
\citealt{1975MNRAS.173..729H}).  Eventually, as the BH--BH binaries
harden, the recoil produced by interactions becomes so great that the
binaries can be ejected from the cluster.  The timescale for merger by
GW emission usually remains longer than the interaction time in the BH
cluster so that hardened binaries are almost always
ejected. Eventually the number of BHs is depleted, and no more than a
few BHs would remain in the cluster core. 
This would imply that BH growth through successive mergers in the
cluster cannot occur. Even if $\sim 1$ BH remained at the center of
every globular cluster today, it is unlikely that this would ever
become detectable as an X-ray binary \citep*{2004ApJ...601L.171K}.

In order to better understand the complex interactions of BHs in
clusters, \citet{2000ApJ...528L..17P} performed direct $N$-body
simulations of systems with $N = 2048$ and $N = 4096$ total stars,
including a small fraction ($\sim1\%$) of equal--mass ``BHs'' 10 times
more massive than the other stars.  They found that $\sim 90\%$ of the
BHs were ejected from the cluster after $4\,-\,10$ relaxation times of
the cluster (less than a few Gyr for most clusters), including $\sim
30\%$ in BH--BH binaries.  Similar $N$-body simulations were most
recently performed by \citet{2004ApJ...608L..25M}, who studied the
formation of a core of BHs in a two--component cluster, with $N =
10^4$.  \citet{2004ApJ...608L..25M} found that the BHs completely
segregate to the core within $\sim 100\,$Myr (ignoring initial mass
segregation of the higher mass progenitors).  These results are
consistent with our qualitative understanding of the interactions of
BHs, and strongly supports the assumptions we will make in \S~\ref{manda}.

Several studies proposed scenarios that could help BH--BH binaries
merge inside clusters, opening the possibility of BH growth and
IMBH formation through successive mergers.
\citet{2002MNRAS.330..232M} suggested that one larger seed BH may help overcome the
Newtonian recoil, since the mass of the binary would be too large to
have a recoil velocity greater than the escape velocity from the
cluster, thus anchoring it in the core. It has also been proposed by
\citet{2002ApJ...576..894M} and \citet{2003ApJ...598..419W} that binary--binary interactions may
have a large influence on BH mergers in the cluster core. These
authors suggest that binary--binary interactions can produce
significant numbers of
long-lived hierarchical triple systems in which the outer BH increases
the inner binary's eccentricity via Kozai-type secular
perturbations \citep{2000ApJ...535..385F}, thereby
increasing the merger rate.  Because these hierarchical triples may be
driven to merge before their next interaction they should have a
higher probability of staying in the cluster, and this can be a
mechanism for retaining merging BHs.

\citet*{2004ApJ...616..221G} (hereafter GMH04) were the first to look at the
possibility of successive mergers of BHs in a cluster core.
In their simulations, GMH04 computed successive interactions between
a BH--BH binary of varying mass ratio with single $10\,M_{\odot}$
BHs sampled from an isotropic background.
Between interactions the binary was evolved according to general
relativity. They concluded that GW emission allowed
for more mergers than previously thought possible, but the number of
BHs required to form an IMBH would be much greater than believed to
exist in a typical globular cluster.

In our simulations, we treat not only binary--single interactions, but
also four-body (binary--binary) interactions. Most importantly, we
compute dynamical interactions for a realistic mixture of single and
binary BHs self-consistently within a ``pure BH'' cluster core. We 
implement a
treatment for the secular evolution of triples, including the Kozai
mechanism \citep{2003ApJ...598..419W,2002ApJ...576..894M}.  We also include, for the first time
in any dynamical treatment, a more realistic IMF for the BHs (to be 
detailed in \S \ref{init}).

Our paper is organized as follows.  In \S \ref{manda} we present our
simulation methods and assumptions, as well as all our initial
conditions. The main results of our simulations are shown in \S
\ref{results}, where we look at evolution of all BH sub-clusters, and in
\S \ref{gravwave}, where we look at merging binary BHs as a source of
GWs.  Finally we conclude our paper in \S \ref{discussion} with a
summary and discussion of the implications of our results, and
suggestions for further studies.

\section{Methods and Assumptions}
\label{manda}
\subsection{Numerical Methods}
\label{methods}
Ideally, to simulate the evolution of $\sim 10^3$ BHs in a massive
cluster, one would like to do a full $N$-body simulation of the
entire cluster, including the BHs and other stars.  Or, capitalizing
on the fact that the BH subsystem dynamically decouples from the rest
of the cluster early on, one could perform an $N$-body simulation of
just the BHs, subject to the cluster potential due to the other stars
and the effects of dynamical friction, which tend to bring BHs that
have been kicked out of the core---but not out of the cluster---back into
the core on a short timescale.  Although the second scenario is
evidently computationally much cheaper than the first to treat via
direct $N$-body techniques, we decided to adopt an even faster
technique which, although approximate, includes all the vital physics
of the problem.  This allows us to sample a wider range of initial
conditions, and make a more thorough map of the parameter space of the
problem.

We treat the evolution of a BH subsystem in a background cluster
subject to binary interactions, using a Monte Carlo technique to
sample interaction rates and treat mass segregation, in conjunction
with the small $N$-body toolkit \Fewbody\ to numerically
integrate binary interactions \citep{2005MNRAS.358..572I}.  We assume that the BH
sub-cluster has a constant density and velocity dispersion throughout
its evolution.  The justification for this assumption is that with any
reasonable initial binary fraction ($\gtrsim$ a few percent), the
sub-cluster will enjoy a long-lived binary-burning phase in which
its core parameters are roughly constant with time \citep{2003ApJ...593..772F}.  The
code we use is a modified version of the one presented in
\citet{2005MNRAS.358..572I}, specially adapted to treat a BH system.

Each dynamical binary interaction is followed using \Fewbody, a
numerical toolkit for simulating small-$N$ gravitational dynamics
that provides automatic calculation termination and classification of
outcomes \citep[for a detailed description see][]{2004MNRAS.352....1F}.
\Fewbody\ numerically integrates the orbits of small-$N$ systems, and
automatically classifies and terminates calculations when an
unambiguous outcome is reached.  Thus it is well-suited for carrying
out large numbers of binary interactions, which can be quite complex
and long-lived, and thus must be treated as computationally
efficiently as possible.

Star clusters are characterized by a dense central core, surrounded by
a much larger low-density halo.  Consistent with this structure of a
star cluster, we assume that all strong interactions occur within the
core, which has a velocity dispersion $\sigma_{\rm core}$ (cf.\
eq.~[\ref{veldispeq}]).  If a product of an interaction has a velocity
greater than the escape velocity from the core of the cluster, $v_{\rm
esc}$, then it is assumed ejected from the cluster and is removed from
the simulation. If it is less than $v_{\rm esc}$ but greater than the
escape velocity from the core into the halo $v_{\rm halo}$, it is
placed in the halo of the cluster, from where it can later reenter the
core through dynamical friction with the background stars. Dynamical
friction is implemented in our code by sampling from a Poisson
distribution with an average timescale given in equation~(\ref{eqseg})
with the average stellar mass, $\left<m\right> = 1.0\, M_{\odot}$
(see \S3.3 from \citealt{2005MNRAS.358..572I}).  
    
Between interactions, all BH--BH binaries are evolved according to the
standard post-Newtonian equations \citep{1964PhRv..136.1224P},
\begin{equation}
\frac{da}{dt} =-\frac{64}{5} \frac{G^3}{c^5}\frac{m_{1}^{2}\,m_{2}^{2}\,(m_{1}+m_{2})}{a^{3}(1-e^{2})^{7/2}}\left(1+\frac{73}{24} e^{2}+\frac{37}{96} e^{4}\right)
\label{pet1}
\end{equation}
\begin{equation}
\frac{de}{dt} = -\frac{304}{15}\frac{G^3}{c^5}\frac{m_{1}\,m_{2}\,(m_{1}+m_{2})e}{a^{4}(1-e^{2})^{5/2}}\left(1+\frac{121}{304}e^{2}\right),
\label{pet2}
\end{equation}
where $m_{1}$ and $m_{2}$ are the masses of the two BHs, $a$ is the
binary semimajor axis, and $e$ is the orbital eccentricity.

In some simulations, we account for linear momentum kicks imparted to
the binary due to the asymmetry of the GW emission \citep{1983MNRAS.203.1049F}.
Because of the large theoretical uncertainty \citep*{2004ApJ...607L...5F, 2005Blanchetetal} in the
recoil velocity of ``major mergers'' (i.e., when the mass ratio $q =
m_2/m_1 \ga 0.4$; here we assume $m_1 > m_2$), and the smaller but
significant uncertainty from the spins of the BHs, we opted to neglect
spin in determining recoil velocities in our simulations. We determine
the overall recoil velocity of the merger remnant, $V_{\rm rec}$, by
using the form of the equation derived by \citet{1983MNRAS.203.1049F},
\begin{equation}
\label{eqn3}
\tilde{V}_{\rm{rec}} = V_{0}\frac{f(q)}{f_{\rm{max}}}\left(\frac{2GM/c^2}{r_{\rm{isco}}}\right)^4,
\end{equation}
where $f(q)=q^2(1-q)/(1+q)^5$, $f_{max} \approx .0179$, $V_{0}$ is the
maximum magnitude of recoil, and $r_{\rm isco}$ is the radius of
innermost stable circular orbit.  \citet{1983MNRAS.203.1049F} found for circular
orbits $V_{0} \approx 1480\,$ km s$^{-1}$, much greater than the
escape velocity from any globular cluster, whereas \citet{2004ApJ...607L...5F} found
$V_0 \sim 10\ -\ 100\,$ km s$^{-1}$.  In our simulations we set
\begin{equation}
V_{\rm{rec}} = V_{0} \frac{ f(q)}{f_{\rm{max}} }
\label{recoil}
\end{equation}
for ease of comparing $V_0$ with other works. We use $V_0$ near or
slightly above the escape velocity of the cluster, up to $80\,$ km
s$^{-1}$.  The form of equation~(\ref{recoil}) is consistent with the
analysis of \citep{2005Blanchetetal}, who found the recoil velocity
for the merger of two non-spinning BHs to be $V_{0} \approx 250 \pm
50\,$km s$^{-1}$ at the second post-Newtonian order. We do not account
for GW recoil in the merger of the inner binaries that are part of
hierarchical triples, because in the simulations where we include GW
recoil, mergers in hierarchical triples are insignificant.

\citet{1993ApJ...418..147L} shows that for the velocity dispersions ( $< 100 $ km
s$^{-1}$) and numbers of BHs ( $ \la 10^3$) expected in the star
clusters we are investigating, the rate of two-body binary formation
from gravitational bremsstrahlung is much less than that of regular
(Newtonian) three-body binary formation, whereby a binary is formed
with the help of a third BH, which takes away the excess energy needed
to form the bound pair.  Therefore, for our simulations, we only
account for three-body and binary--binary (four-body) Newtonian
interactions.

In a dense sub-cluster of BHs, three-body binary formation can lead
to the formation of a significant number of BH binaries, and
eventually can help lead to the disruption of the entire
sub-cluster. \citet{2005MNRAS.358..572I} calculated the three-body binary formation
rate for a binary of minimum hardness 
\begin{equation}
\eta_{\rm min} = \frac{G\,m_1\,m_2}{b_{\rm max} \left<m\right> \sigma_{\rm core}^2},
\end{equation}
where $b_{\rm max}$ is the maximum size of the region in which the three
objects interact and $\sigma_{\rm core}$ is the three-dimensional velocity
dispersion of the core.  The final rate per star they found for the
formation of a binary with hardness $\eta > \eta_{\rm min}$ is
\begin{equation}
\Gamma(\eta > \eta_{\rm min}) = \pi \frac{n^2_{\rm c}\,G^5 \left<m\right>^5}{\sigma_{\rm core}^9} f(m_1,m_2,m_3,\eta)
\label{threebody}
\end{equation}
where
\begin{eqnarray}
f(m_1,m_2,m_3,\eta) = \frac{n_2\,n_3}{n^2_{\rm c}} \frac{m^5_1}{\left<m\right>^5} \frac{m^5_2}{\left<m\right>^5} \eta^{-5}(1+2\eta) \nonumber \\
\times\left(1+\frac{v_3}{\sigma_{\rm core}} \eta^{-1/2}\sqrt{2\frac{m_1\,m_2}{(m1+m2)\left<m\right>}}\right)\frac{v_{12}}{\sigma_{\rm core}},
\end{eqnarray}
$n_c$ is the core density of the BHs, $n_2$ is the core density of
objects of mass $m_2$, $n_3$ is the core density of objects of mass
$m_3$, $v_{12}$ is the relative velocity of the first object with
respect to the second, and $v_3$ is the relative velocity of the third
object with respect to the center of mass of the first two objects.
We follow the exact treatment of dynamical interactions of
\citet{2005MNRAS.358..572I} but include three-body binary formation
with $\eta_{\rm min} = 1$ in a consistent manner (see, in particular,
their \S 3.4).  We note that a more accurate criterion for the minimum binding
energy we use should be based on the orbital velocity of the lightest
member of the formed binary since we are not looking at equal--mass
clusters \citep{1990AJ.....99..979H}. However, since we typically do
not have mass ratios above $\sim 10$, we find our criterion to be sufficient.

Because the code is not yet capable of following the evolution of
triples between interactions, we must break them up before the next
interaction time-step. In order to determine how to destroy the
triple, we check if the binary is likely to merge before its next
interaction.  As a first step, we integrate numerically
equations~(\ref{pet1})~and~(\ref{pet2}) \citep{1964PhRv..136.1224P}.
We also scale the timescale of merger according to
\citet{2002ApJ...576..894M} by calculating the maximum eccentricity
from a first order Kozai mechanism approximation without
post-Newtonian precession by solving equation~(8) of
\citet{2003ApJ...598..419W}. We then use the smaller merger time of
the two methods. It is necessary to consider both methods because the
scaling from \citet{2002ApJ...576..894M} overestimates the merging
time in the instances when the Kozai mechanism is insignificant.  In
this case, the inner binary is merely perturbed and the eccentricity
does not fluctuate, therefore the timescale of the merger should be
the same as for an unperturbed binary.  If the inner binary is likely
to merge before the triple would interact with a field BH or BH--BH
binary it is immediately merged, otherwise the triple is broken-up
keeping the inner binary. Here, we assume that the outer BH is ejected
from the triple with a velocity equal to its relative velocity with
the center of mass of the triple. We keep the inner binary, but shrink
the binary separation to conserve the energy of the system.

\subsection{Initial Conditions and Parameters}
\label{init}

\begin{deluxetable*}{lccccccccccc}
\tabletypesize{\scriptsize}
\notetoeditor{This table should not be rotated in final manuscript, and should be located near the Initial Conditions and Parameters section (ref -- init)}
\tablecolumns{12}
\tablewidth{0pc}
\tablecaption{Simulation Parameters}
\tablehead{
\colhead{}&
\colhead{Structure}&
\colhead{$M_{\rm cl}$}&
\colhead{}&
\colhead{}&
\colhead{$n_c$}&
\colhead{$t_{\rm rh}$}&
\colhead{$\sigma_{\rm 1,core}$}&
\colhead{$\sigma_{\rm 1,BH}$}&
\colhead{$v_{\rm esc}$}&
\colhead{$v_{\rm halo,esc}$}&
\colhead{$v_{\rm halo}$}
\\
\colhead{Model Name}&
\colhead{($W_0$)}&
\colhead{($M_{\odot}$)}&
\colhead{Effective $N$}&
\colhead{$N_{\rm BH}$}&
\colhead{(pc$^{-3}$)}&
\colhead{(yr)}&
\colhead{(km s$^{-1}$)}&
\colhead{(km s$^{-1}$)}&
\colhead{(km s$^{-1}$)}&
\colhead{(km s$^{-1}$)}&
\colhead{(km s$^{-1}$)}
}
\startdata
e5e5king7\dotfill      &       7    &   5$\times 10^5$   &   1$\times 10^6$   &    512   &   5$\times 10^5$   &   1.5$\times 10^8$   &   14.1   &   14.1$\phn$   &   55.6   &   38.0   &   40.7\\
v2e5k7\dotfill         &       7    &   5$\times 10^5$   &   1$\times 10^6$   &    512   &   5$\times 10^5$   &   1.5$\times 10^8$   &   14.1   &   7.0   &   55.6   &   38.0   &   40.7\\
v3e5k7\dotfill         &       7    &   5$\times 10^5$   &   1$\times 10^6$   &    512   &   5$\times 10^5$   &   1.5$\times 10^8$   &   14.1   &   4.7   &   55.6   &   38.0   &   40.7\\
v3e5k7ej54\tablenotemark{a}\dotfill     &       7    &   5$\times 10^5$   &   1$\times 10^6$   &    512   &   5$\times 10^5$   &   1.5$\times 10^8$   &   14.1   &   4.7   &   55.6   &   38.0   &   40.7\\
v3e5k7ej75\tablenotemark{a}\dotfill     &       7    &   5$\times 10^5$   &   1$\times 10^6$   &    512   &   5$\times 10^5$   &   1.5$\times 10^8$   &   14.1   &   4.7   &   55.6   &   38.0   &   40.7\\
v5e5k7\dotfill        &       7    &   5$\times 10^5$   &   1$\times 10^6$   &    512   &   5$\times 10^5$   &   1.5$\times 10^8$   &   14.1   &   2.8   &   55.6   &   38.0   &   40.7\\
e5e5king9\dotfill      &       9    &   5$\times 10^5$   &   1$\times 10^6$   &    512   &   1$\times 10^5$   &   7.1$\times 10^8$   &   10.1   &   10.1$\phn$   &   44.1   &   22.7   &   38.0\\
v2e5k9\dotfill         &       9    &   5$\times 10^5$   &   1$\times 10^6$   &    512   &   1$\times 10^5$   &   7.1$\times 10^8$   &   10.1   &   5.1   &   44.1   &   22.7   &   38.0\\
v3e5k9\dotfill         &       9    &   5$\times 10^5$   &   1$\times 10^6$   &    512   &   1$\times 10^5$   &   7.1$\times 10^8$   &   10.1   &   3.4   &   44.1   &   22.7   &   38.0\\
e55king9\dotfill       &       9    &   5$\times 10^5$   &   1$\times 10^6$   &    512   &   5$\times 10^5$   &   3.2$\times 10^8$   &   13.2   &   13.2$\phn$   &   57.6   &   29.6   &   49.5\\
v2e55k9--256\dotfill    &       9    &   5$\times 10^5$   &   1$\times 10^6$   &    256   &   5$\times 10^5$   &   3.2$\times 10^8$   &   13.2   &   6.6   &   57.6   &   29.6   &   49.5\\
v2e55k9\dotfill        &       9    &   5$\times 10^5$   &   1$\times 10^6$   &    512   &   5$\times 10^5$   &   3.2$\times 10^8$   &   13.2   &   6.6   &   57.6   &   29.6   &   49.5\\
v2e55k9--1024\dotfill   &       9    &   5$\times 10^5$   &   1$\times 10^6$   &   1024$\phn$   &   5$\times 10^5$   &   3.2$\times 10^8$   &   13.2   &   6.6   &   57.6   &   29.6   &   49.5\\
v2e55k9--2048\dotfill   &       9    &   5$\times 10^5$   &   1$\times 10^6$   &   2048$\phn$   &   5$\times 10^5$   &   3.2$\times 10^8$   &   13.2   &   6.6   &   57.6   &   29.6   &   49.5\\
v2e55k9--100\tablenotemark{b}\dotfill    &       9    &   5$\times 10^5$   &   1$\times 10^6$   &    513   &   5$\times 10^5$   &   3.2$\times 10^8$   &   13.2   &   6.6   &   57.6   &   29.6   &   49.5\\
v2e55k9--200\tablenotemark{b}\dotfill    &       9    &   5$\times 10^5$   &   1$\times 10^6$   &    513   &   5$\times 10^5$   &   3.2$\times 10^8$   &   13.2   &   6.6   &   57.6   &   29.6   &   49.5\\
v2e55k9e6\tablenotemark{a}\dotfill     &       9    &   5$\times 10^5$   &   1$\times 10^6$   &    512   &   5$\times 10^5$   &   3.2$\times 10^8$   &   13.2   &   6.6   &   57.6   &   29.6   &   49.5\\
v2e55k9e65\tablenotemark{a}\dotfill     &       9    &   5$\times 10^5$   &   1$\times 10^6$   &    512   &   5$\times 10^5$   &   3.2$\times 10^8$   &   13.2   &   6.6   &   57.6   &   29.6   &   49.5\\
v2e55k9e7\tablenotemark{a}\dotfill      &       9    &   5$\times 10^5$   &   1$\times 10^6$   &    512   &   5$\times 10^5$   &   3.2$\times 10^8$   &   13.2   &   6.6   &   57.6   &   29.6   &   49.5\\
v2e55k9e8\tablenotemark{a}\dotfill      &       9    &   5$\times 10^5$   &   1$\times 10^6$   &    512   &   5$\times 10^5$   &   3.2$\times 10^8$   &   13.2   &   6.6   &   57.6   &   29.6   &   49.5\\
v3e55k9\dotfill        &       9    &   5$\times 10^5$   &   1$\times 10^6$   &    512   &   5$\times 10^5$   &   3.2$\times 10^8$   &   13.2   &   4.4   &   57.6   &   29.6   &   49.5\\
v2e6e5k9\dotfill       &       9    &   1$\times 10^6$   &   2$\times 10^6$   &    512   &   5$\times 10^5$   &   5.9$\times 10^8$   &   16.6   &   8.3   &   72.3   &   37.2   &   62.2\\
v3e6e5k9\dotfill       &       9    &   1$\times 10^6$   &   2$\times 10^6$   &    512   &   5$\times 10^5$   &   5.9$\times 10^8$   &   16.6   &   5.6   &   72.3   &   37.2   &   62.2\\
v22e6e5k9 \dotfill     &       9    &   2$\times 10^6$   &   4$\times 10^6$   &    512   &   5$\times 10^5$   &   1.1$\times 10^9$   &   21.0   &   10.5$\phn$   &   91.3   &   46.9   &   78.6\\
v32e6e5k9\dotfill      &       9    &   2$\times 10^6$   &   4$\times 10^6$   &    512   &   5$\times 10^5$   &   1.1$\times 10^9$   &   21.0   &   7.0     &   91.3   &   46.9   &   78.6\\
e5king9\dotfill        &       9    &   5$\times 10^5$   &   1$\times 10^6$   &    512   &   1$\times 10^6$   &   2.2$\times 10^8$   &   14.9   &   14.9$\phn$   &   64.7   &   33.7   &   55.7\\
v2o5k9\dotfill        &       9    &   5$\times 10^5$   &   1$\times 10^6$   &    512   &   1$\times 10^6$   &   2.2$\times 10^8$   &   14.9   &   7.4   &   64.7   &   33.3   &   55.7\\
v3o5k9\dotfill         &       9    &   5$\times 10^5$   &   1$\times 10^6$   &    512   &   1$\times 10^6$   &   2.2$\times 10^8$   &   14.9   &   5.0   &   64.7   &   33.3   &   55.7\\
e5e5king11\dotfill     &       11$\phn$   &   5$\times 10^5$   &   1$\times 10^6$   &    512   &   1$\times 10^5$   &   2.6$\times 10^9$   &   $\phn$6.9   &   6.9   &   33.3   &   14.8   &   29.7\\
v2e5k11\dotfill        &       11$\phn$   &   5$\times 10^5$   &   1$\times 10^6$   &    512   &   1$\times 10^5$   &   2.6$\times 10^9$   &   $\phn$6.9   &   3.5   &   33.3   &   14.8   &   29.7\\
v3e5k11\dotfill        &       11$\phn$   &   5$\times 10^5$   &   1$\times 10^6$   &    512   &   1$\times 10^5$   &   2.6$\times 10^9$   &   $\phn$6.9   &   2.3   &   33.3   &   14.8   &   29.7\\
e55king11\dotfill      &       11$\phn$   &   5$\times 10^5$   &   1$\times 10^6$   &    512   &   5$\times 10^5$   &   1.2$\times 10^9$   &   $\phn$9.1   &   9.1   &   43.5   &   19.3   &   38.8\\
v2e55k11\dotfill       &       11$\phn$   &   5$\times 10^5$   &   1$\times 10^6$   &    512   &   5$\times 10^5$   &   1.2$\times 10^9$   &   $\phn$9.1   &   4.5   &   43.5   &   19.3   &   38.8\\
v3e55k11\dotfill       &       11$\phn$   &   5$\times 10^5$   &   1$\times 10^6$   &    512   &   5$\times 10^5$   &   1.2$\times 10^9$   &   $\phn$9.1   &   3.0   &   43.5   &   19.3   &   38.8\\
e5king11\dotfill       &       11$\phn$   &   5$\times 10^5$   &   1$\times 10^6$   &    512   &   1$\times 10^6$   &   8.3$\times 10^8$   &   10.2   &   10.2$\phn$   &   48.9   &   21.7   &   43.6\\
v2o5k11\dotfill        &       11$\phn$   &   5$\times 10^5$   &   1$\times 10^6$   &    512   &   1$\times 10^6$   &   8.3$\times 10^8$   &   10.2   &   5.1   &   48.9   &   21.7   &   43.6\\
v2o5k1110\dotfill      &       11$\phn$   &   5$\times 10^5$   &   1$\times 10^6$   &   1024$\phn$   &   1$\times 10^6$   &   8.3$\times 10^8$   &   10.2   &   5.1   &   48.9   &   21.7   &   43.6\\
v2o5k1111\dotfill      &       11$\phn$   &   5$\times 10^5$   &   1$\times 10^6$   &   2048$\phn$   &   1$\times 10^6$   &   8.3$\times 10^8$   &   10.2   &   5.1   &   48.9   &   21.7   &   43.6\\
v3o5k11\dotfill        &       11$\phn$   &   5$\times 10^5$   &   1$\times 10^6$   &    512   &   1$\times 10^6$   &   8.3$\times 10^8$   &   10.2   &   3.4   &   48.9   &   21.7   &   43.6\\
v2e5e7k11\dotfill      &       11$\phn$   &   5$\times 10^5$   &   1$\times 10^6$   &    512   &   1$\times 10^7$   &   2.6$\times 10^8$   &   15.0   &   7.5   &   71.7   &   31.8   &   64.0\\
v3e5e7k11\dotfill      &       11$\phn$   &   5$\times 10^5$   &   1$\times 10^6$   &    512   &   1$\times 10^7$   &   2.6$\times 10^8$   &   15.0   &   5.0   &   71.7   &   31.8   &   64.0\\
GMH\tablenotemark{c}\dotfill &   \nodata$\phn$           &    \nodata         &  \nodata           &    512   &   5$\times 10^5$   &  \nodata             &   \nodata   &   5.8  &   50.0   &   \nodata   &   \nodata \\
\enddata
\tablecomments{ Starting from the top, the table is sorted according to the following parameters in their respective order: $W_0$, $n_c$, $\sigma_{\rm{1,BH}}$, $M_{\rm{cl}}$, $N_{\rm{BH}}$.}
\tablenotetext{a}{These models include GW recoil, as described in \S \ref{gravrad}. Models v3e5k7ej54 and v3e5k7ej75 have maximum recoil velocities $V_0 = 54\,$ and $75\,$ km s$^{-1}$ respectively.  Models v2e55k9e6, v2e55k9e65, v2e55k9e7, and v2e55k9e8 correspond to $V_0 = 60,\ 65,\ 70,\,$ and $80\,$ km s$^{-1}$ }
\tablenotetext{b}{These models include one primordial seed BH of mass
  $100\,M_{\odot}$ and $200\,M_{\odot}$, respectively, as
  detailed in \S \ref{mechs}.}
\tablenotetext{c}{The GMH model parameters correspond to the initial
  conditions used in GMH04.}
\label{tab1}
\end{deluxetable*}

\begin{figure*}
\epsscale{.80}
\plotone{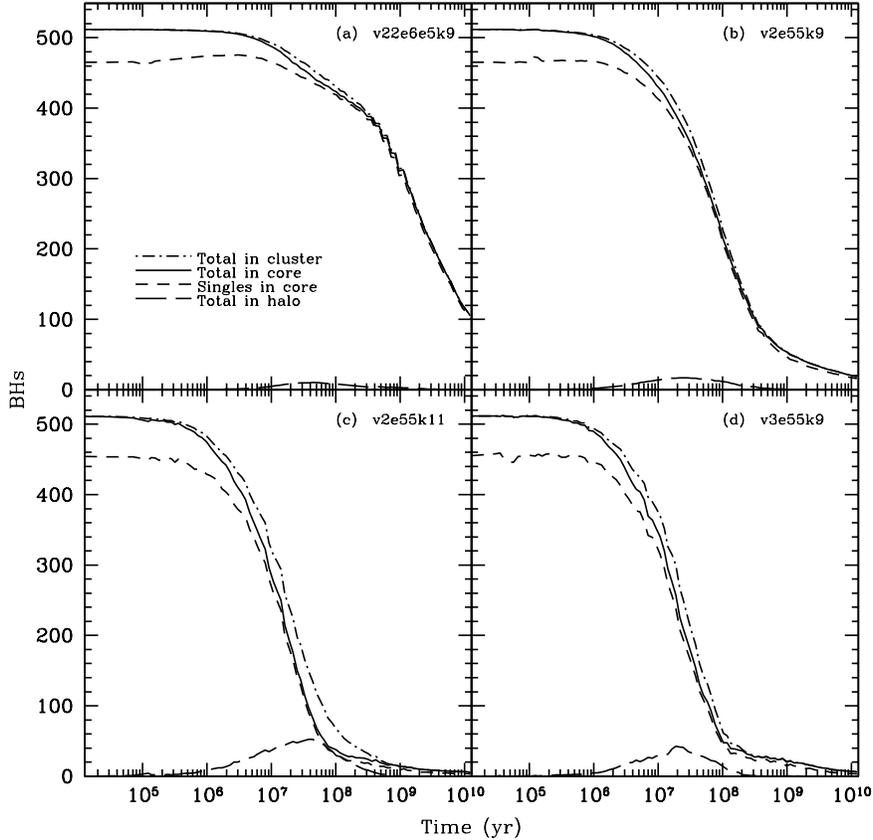}
\caption{Sub-cluster evolution. The data for each
panel are binned and then averaged over several simulations in order
to reduce the level of noise in the graph. The clusters reach
equipartition when the total number of BHs in the core
(\emph{solid}) is $\la 40$. Notice how few BHs are in
the cluster halo (\emph{long-dashed}) throughout the evolution. Only in v2e55k11, panel (c), do the numbers of
BHs in the core and halo become comparable, but only when the cluster
has already approximately reached equipartition.  The more massive
cluster v22e6e5k9, panel (a), does not reach equipartition in a Hubble
time, suggesting massive clusters with low $W_0$ could still contain
significant numbers of single BHs in their cores. Each model has the same core
density, $n_c = 5 \times 10^5\,$pc$^{-3}$, and $K = 4$, except model v3e55k9,
panel (d), which has $K = 1.77$. Model v22e6e5k9 is a massive $W_0 =
9$ King model  with  $M_{\rm cl} = 2 \times 10^6\,M_{\odot}$.  Model v2e55k11 is a $W_0 = 11$ King model with 
$M_{\rm cl} = 5 \times 10^5\,M_{\odot}$. Models v2e55k9, panel
(b), and v3e55k9 are the same as v2e55k11, except they have $W_0 = 9$.
\label{compnum}}
\end{figure*}

We use the results of \citet*{2004ApJ...611.1068B} (hereafter BSR04), adopting the
mass and binary distributions of their standard model at $11.0\,$Myr for
our initial conditions (see, e.g., their Fig. 2, Fig. 4, \&
Fig. 5). BSR04 used a population synthesis approach to follow the
evolution of a large number of massive stars and binaries, as would
likely form in a massive star cluster.  The model we base our
calculations on has an initial binary fraction $f_{\rm{bin}} = 50 \%$,
and follows the traditional Salpeter IMF for all initial stars with
mass $ > 4\,M_{\odot}$. The binary fraction we use for our
simulations is, of course, the final binary fraction found in BSR04,
$f_{\rm{bin}} \approx 14 \%$.  Although most BH binaries have MS
companions, we assume that these binaries will eventually become
BH--BH binaries through exchange interactions. We create BH--BH
binaries in their place, and select the companion BH, such that the
distribution of the mass ratio, q, is uniform throughout the range
$q_{\rm min} < q < 1$ ($q_{\rm min}$ is set by the minimum BH mass in
our distribution), consistent with observations for $q \ga 0.2$
\citep*{2001A&A...376..982W}.  We then increase the separation of the BH--BH binary
assuming the binary preserves its binding energy in the exchange
interaction.  All wide binaries with orbital period $P > 10^{4}$ days
are destroyed before our simulations begin. For each individual run,
the mass of each BH is randomly selected with a distribution that
reflects the results of BSR04, so that no two runs of any model
contain the exact same BH population.

The parameters used in all our simulations can be found in Table
\ref{tab1}.  For our simulations, aside from the exceptions noted in
the table, we use self-consistent parameters determined by a King
model, with $W_0 = 7$, 9, and 11.  Given a total cluster mass, $M_{\rm
cl}$, and core density, $n_{\rm c}$, we can calculate the one-dimensional
velocity dispersion, $\sigma_{\rm 1,core}$, the escape velocity from
the center of the potential to the half-mass radius, $v_{\rm halo}$,
the escape velocity of a BH in the core from the entire cluster,
$v_{\rm esc}$, and the escape velocity of a BH at the half-mass
radius from the entire cluster, $v_{\rm halo,esc}$. We analyzed
relatively massive and dense clusters, precisely the types of clusters
where BH growth would be expected. Specifically, we systematically
varied $M_{\rm cl}$ between $5\times 10^5\, $ and $2\times 10^6
\rm{M_{\odot}}$, and $n_c$ between $ 10^5\, $ and $10^7 \rm{pc}^{-3}$.
Most of our simulations had 512 BHs ($N_{\rm BH}$ = 512), but in two
simulations we looked at clusters with smaller and larger numbers of
BHs.  All of these parameters are listed in Table~\ref{tab1}.  The escape velocities are used to determine whether the
product of an interaction is to remain in the cluster as prescribed in
\S \ref{methods}.  The velocity dispersion of the BHs, $\sigma_{\rm
BH}$, can be related to the one-dimensional velocity dispersion simply as
$\sigma_{\rm BH} = \sqrt{3} \sigma_{\rm 1,BH}$.

The properties of the decoupled sub-cluster in relation to the
properties of the cluster core are not so apparent.  In our
simulations we need to know the initial velocity dispersion of the
BHs, $\sigma_{\rm BH}$. Numerical simulations suggest that the mean
kinetic energy of the dynamically decoupled massive objects is only a
few times larger than for other objects in the core
\citep{2004ApJ...604..632G}. When decoupling, the BHs will have an effective
velocity dispersion which we write as
\begin{equation}
\sigma_{\rm BH} = \left(K\frac{\langle m\rangle}{\langle M_{\rm BH}\rangle}\right)^{1/2}\sigma_{\rm core},
\label{veldispeq}
\end{equation}
where $\sigma_{\rm{core}}$ is the velocity dispersion of the core, $
\left<M_{\rm BH}\right>$ is the average BH mass, and $K$ is the ratio
of the mean kinetic energy of the BHs to that of the rest of the core,
with $K = 1$ corresponding to complete energy equipartition. Because
the BHs are dynamically decoupled from the rest of the cluster, they
will undergo their own independent evolution as a small cluster of
stars, increasing their density and velocity dispersion in the
process.  Therefore we look at sub-clusters with $K = .64$, $1.77$,
$4$, and $16$.

\section{Cluster Dynamics and BH Growth}
\label{results}
In this section, we discuss our results for the evolution of the BH sub-cluster and we
analyze the conditions under which successive mergers of BHs  lead to
the growth and retention of candidate IMBHs. 

In most of our simulations the BH sub-clusters reach equipartition
(our standard for the complete disruption of the sub-cluster) in a few
Gyr. In the most extreme cases the sub-cluster may dissipate in less
than $100\,$Myr, as in the very dense model v3e5e7k11; or not at all,
as in all models with $K = 16$.  We also find there is a significant
probability a BH with mass $\ga 100\,M_{\odot}$ can form. For clusters
that reach equipartition, massive BHs are most likely to form in clusters with 
high core densities and temperatures.  This also leads to more growth
for a given King profile. Also, King models with lower
$W_0$ are more likely to form candidate IMBHs.  The growth of massive
BHs is highly dependent on the maximum GW recoil velocity, nearly
stopping it completely even when it is only a few percent larger than the core
escape speed.

\subsection{Fate of the BHs}
\label{bhfate}

\begin{figure*}
\epsscale{1.0}
\plottwo{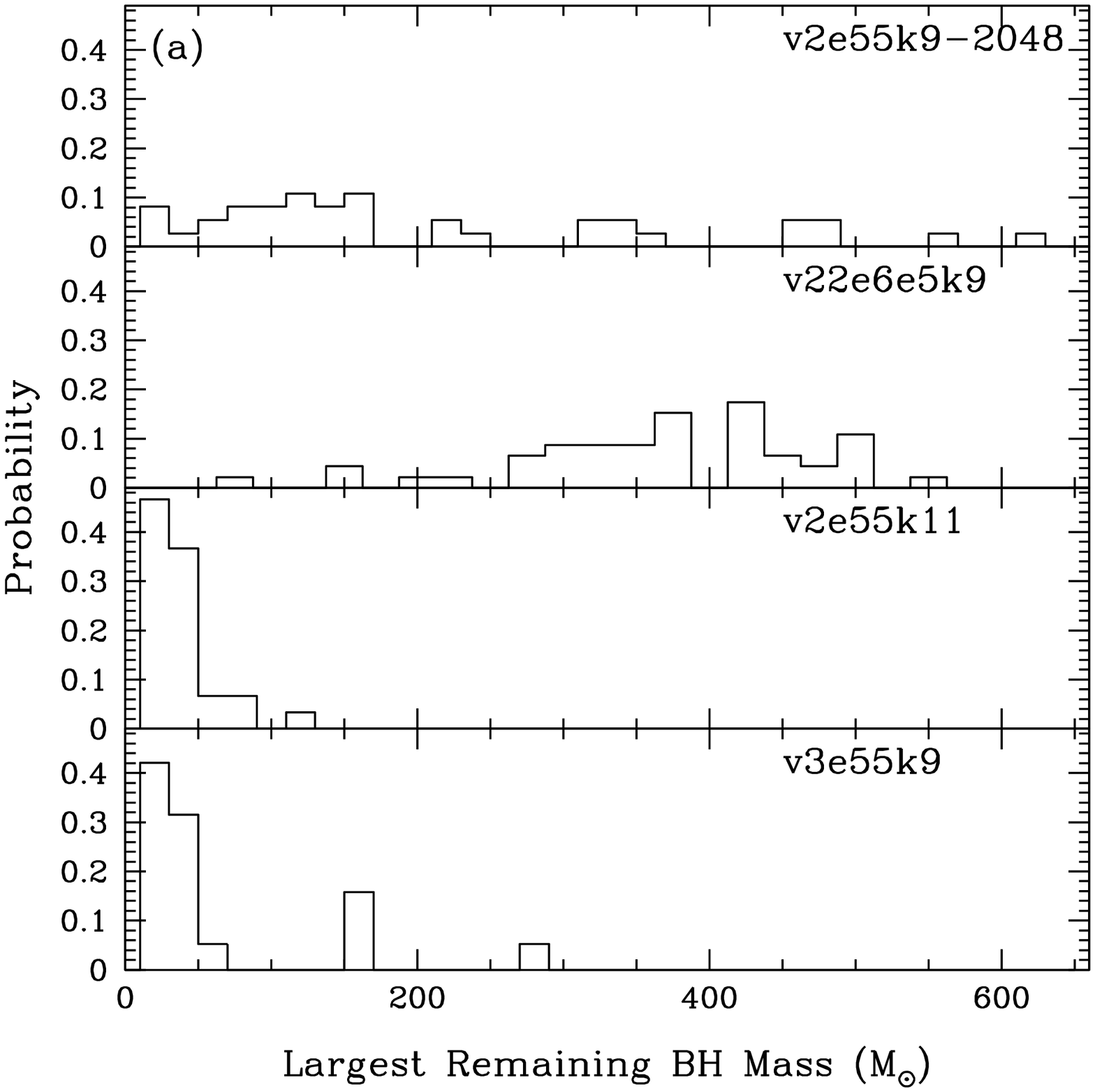}{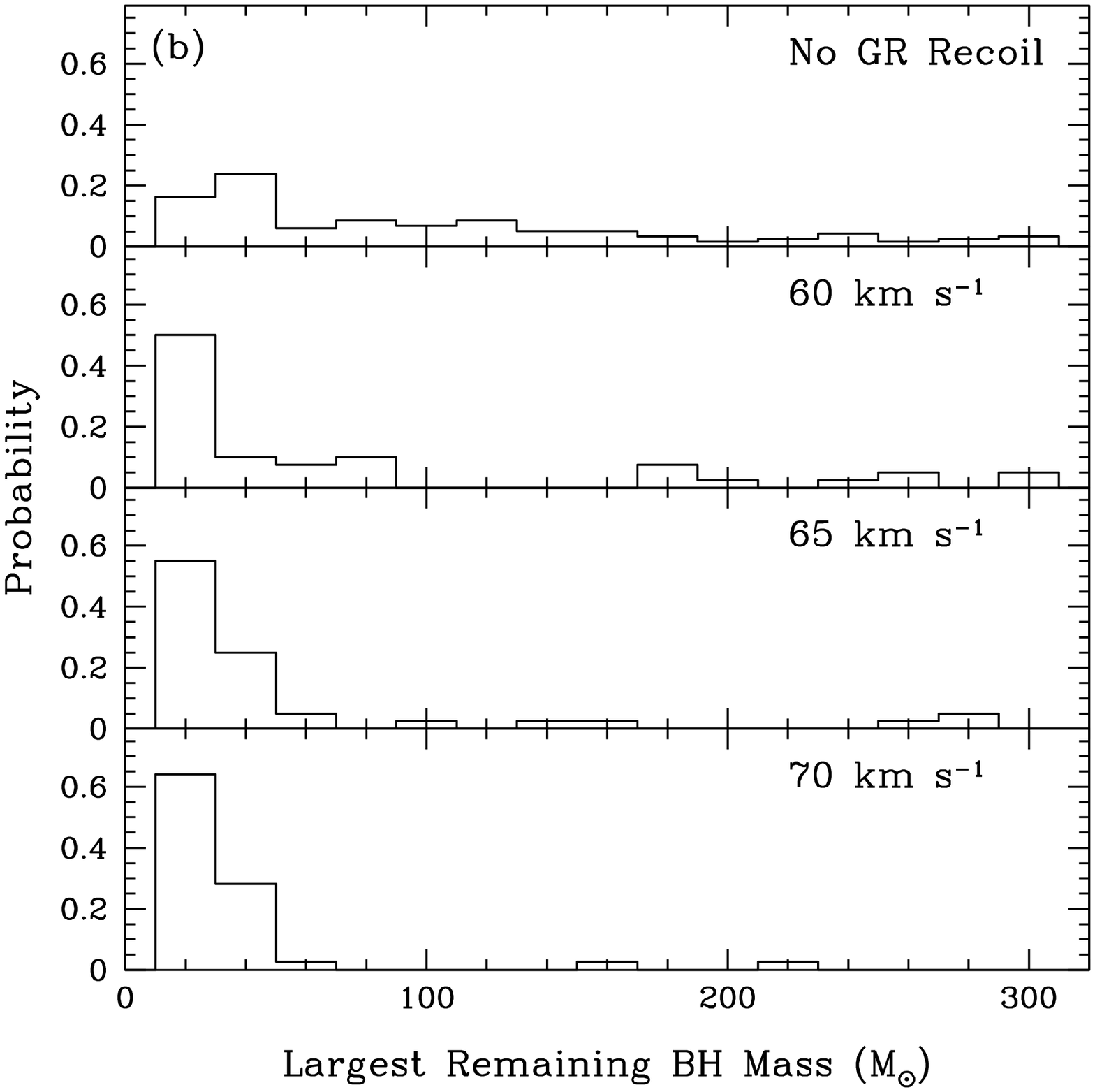}
\caption{Mass distribution of largest remaining BH. The largest remaining BH
is simply the most massive BH remaining  in
the cluster at equipartition.  Panel~(a)
is the mass distribution from a few different cluster types, with the model name in the upper right-hand corner.  Panel~(b) is
the mass distribution from cluster model v2e55k9, with varying GW
recoil kick velocities, $V_0$, corresponding to, from the top of the figure down, models
v2e55k9, v2e55k9e6, v2e55k9e65, and v2e55k9e7.  }
\label{masseject}
\end{figure*}

The fate of the BHs in the cluster is determined mainly by their
characteristic interaction rate in the core.  However, two main
questions remain regarding how the space of initial cluster parameters
is divided among the different possible outcomes: does the BH
sub-cluster reach equipartition with the cluster stars within a
Hubble time? Do the BHs experience enough successive mergers
to form an IMBH?  Because of the simplicity of our model and speed of
our code, the trends associated with varying one parameter in the
simulation are evident.  

Figure~\ref{compnum} shows the number of BHs located in the
cluster core and halo as a function of time for a variety of different
cluster types. We see that most BHs remain in the cluster core rather
than in the halo of the cluster. A fraction $\la 3\%$ of all the BH--BH mergers from cluster occur within the
halo. The sudden drop in the binary fraction, as seen again in
Figure~\ref{compnum}, can be attributed to strong binary--binary
interactions. Even though this causes a low binary fraction for the
majority of the dynamical evolution of the sub-cluster, approximately $10
- 15\%$ of the BHs are ejected in binaries when the BH
sub-cluster reaches equipartition in a Hubble time.

For sub-clusters with low densities and high velocity dispersions, the
three-body binary formation rate can become so small that a newly formed
binary is disrupted before the next binary formation.  Despite this,
our simulations suggest that there always exists at least one binary
in the core that may be able to keep the sub-cluster from undergoing
core collapse. This binary, among others, is created through
three-body binary formation, which is the underlying mechanism for the
entire evolution of the sub-cluster.  Since BHs are only ejected from
the cluster after three- or four-body interactions, BH sub-clusters
that do not produce binaries at a high rate do not dissolve within a
Hubble time (see, e.g., models e5e5king7, e5king9, and v22e6e5k9). If
the sub-clusters' parameters have not changed significantly over the
evolution of the entire cluster, then a significant number of single BHs could
exist in clusters similar to those considered in our simulations.

\subsection{Formation of an IMBH}
\label{formimbh}

Our simulations uniquely allow us to follow the interactions of a
realistic mix of single-- and binary--BHs and monitor the growth of BHs
through successive mergers.  
We find that in a given King model, clusters with greater core
densities and larger masses have a greater probability of forming an
IMBH.  They usually grow to even larger masses.  For a fixed mass
cluster that fits a given King model profile larger core densities
result in a smaller half-mass relaxation time, $t_{\rm{rh}}$.  If this
timescale is small enough, the cluster may actually undergo runaway
growth through stellar collisions.  This suggests that one is more
likely to find an IMBH, whether formed through successive BH mergers
or stellar collisions, in clusters with dense cores.

However, clusters with lower degrees of central concentration
(smaller $W_0$, and higher $\sigma_{\rm core}$) seem to have more BH growth. Figure
\ref{masseject} shows the mass distributions of the largest BHs formed
in a few different cluster types. We see that when clusters completely
disrupt in less than a Hubble time, larger core temperatures directly
correlate with more growth of a single massive BH.  

This is evident in the directly comparing models v2e55k9 and v3e55k9
(where $K = 4,\,$ and $1.77$ respectively). Although the difference in
the average number of mergers in the cluster is small---v2e55k9 had
about 14 mergers whereas v3e55k9 had about 13---the number and sizes
of the large BHs formed are dramatically different.  Model v2e55k9 had
about twice the number of BHs as v3e55k9 with mass $> 100\,M_{\odot}$
remain in the core at equipartition, and an average mass of the
largest BH remaining in the core about $60\%$ larger as well. In model
v3e55k9, the velocity dispersion of the BHs, $\sigma_{\rm BH}$, is
lower than v2e55k9.  As is evident in equation~(\ref{threebody}), the
interaction rate increases rapidly with lower velocity dispersions,
and hence leads to a quicker dissipation of the BH sub-cluster.  This
is consistent with our understanding that three-body binary formation
drives the evolution to the eventual equipartition of the
system.

In clusters where there are successive BH mergers, the first formation
of BHs with mass $\ga 100\,M_{\odot}$ occurs at $\sim 10\,$Myr after
the sub-cluster formed, roughly independent of the cluster model. The
probability that the cluster has a BH with mass $\ga 100\,M_{\odot}$ is then
proportional to the logarithm of time (until the cluster reaches
equipartition). On such a short timescale, one may worry that the BHs
could also be colliding with massive stars, counter to our assumption
of a "pure BH" system. However, as we discussed in \S~\ref{intro}, for
clusters in which runaway collisions of massive stars were avoided,
the massive stars are not expected to enter the BH core before
exploding as supernovae.

\subsection{Proposed Mechanisms for BH Retention}
\label{mechs}
One explanation for why BHs would stay in  the cluster and not be
ejected by hardening and eventual recoil is the Kozai
mechanism in stable hierarchical triples \citep{2002ApJ...576..894M,2003ApJ...598..419W}. If the
relative inclination between the orbital planes in a given triple is large enough,
the inner binary's eccentricity can be driven to a high value
(formally to 1).  The high eccentricity necessarily means a decrease
in the merging time of the
inner binary so that the inner binary will merge before the triple's next likely
interaction. We find that, typically, this has no significant
effect on the formation of large BHs.  
 Overall, mergers caused by the Kozai mechanism
account for less than $\approx 10\%$ of the total mergers, and a
negligible amount in most models.  For example, the model with the
largest percentage of triple mergers, v3e5k11, exhibits almost no
growth at all, with none of the 14 runs having a BH with mass greater
than even $80\,M_{\odot}$.  

Another possibility that favors the retention of large
BHs in clusters is the introduction of a massive seed BH
\citep{2002MNRAS.330..232M}. The mass spectrum of BHs given in BSR04
generally includes a BH of $\sim 50\,M_{\odot}$ in each cluster. The
presence of a BH of this mass does not always mean that this BH will remain
in the cluster. In fact, the largest initial BH is almost always
ejected from the cluster. Although \citet{2002MNRAS.330..232M} suggest
that the introduction of a seed BH with mass $50\,M_{\odot}$ may be
sufficient to cause significant growth, we find that this is still not
massive enough, in agreement with GMH04.
 
In model v2e55k9--100 we introduce an initial seed BH of mass
$100\,M_{\odot}$; in model v2e55k9--200 a $200\,M_{\odot}$ seed BH.
We find that even BHs with these large masses can easily be ejected
from the cluster.  For example, in model v2e55k9--100, the seed BH was
retained only $18\%$ of the time.  Even in model v2e55k9--200, with a
seed BH of $200\ {\rm M_{\odot}}$, the seed BH is again retained only
$35\%$ of the time. These probabilities are still lower than those in
GMH04, where they found the BHs to be retained in a similar cluster
$\sim 40\%$ and $\sim 90\%$ of the time, respectively. This
discrepancy can likely be attributed to the mass distribution of BHs
in our simulations.  The average mass of the BHs in our simulation is
$50\%$ higher than in GMH04, and thus there is an increased
probability that the large seed BHs will be ejected.

\subsection{GW Recoil}
\label{gravrad}
One key question regarding coalescing binary BHs
is the magnitude of linear recoil caused by the asymmetry of the GW emitted by
the binary.  Our code allows us to prescribe a systemic recoil
velocity for every BH--BH merger.  Through this we can follow the
effects of GW recoil in a cluster environment. We are able to
understand the possible effects of gravitational recoil by varying the
maximum magnitude of the recoil velocity in the base model v2e55k9
\citep{2004ApJ...607L...5F, 2005Blanchetetal}.

As expected, the number of successive mergers has a strong dependence
on the maximum recoil velocity; however, it seems that it has only a
small effect on the overall dynamics of the rest of the BH
sub-cluster. Increasing the recoil has almost no noticeable effect on
the total number of mergers in the cluster, but can have significant
consequences on the rate of BH--BH inspirals (see \S
\ref{gravwave}). Even a maximum recoil velocity slightly larger than
the escape velocity of the core ($V_{0} = 1.042 \times v_{\rm{esc}} =
60$ km s$^{-1}$) dramatically changes the number of large BHs formed in
this model, as seen in Figure~\ref{masseject}.  With this recoil
velocity, the probability of forming a $100\ {\rm M_{\odot}}$ is cut
in half. When we look at higher recoil velocities, the
possibility of growing a large BH gets only smaller.

This is not such a surprising result when one considers our
simple prescription for modeling GW recoil. In mergers with a
mass ratio, $q = m_2/ m_1 \approx~.38$, or when $f(q) \approx
f_{max}$, the coalescing binary will generally be ejected when $V_{0}$
is close to the core escape speed (see eq.~[\ref{eqn3}]).  For BHs
which go through successive mergers, it is likely that before
reaching masses $> 100\,M_{\odot}$ the BH will have already
been ejected. However, since this only affects BHs after they merge,
the number of mergers remains relatively unaffected.

Our simple model of GW recoil neglects some aspects of the process,
which could alter results.  In particular, we do
not follow the evolution of the BHs' spins.  Because of this, we must
neglect the effects of spin on recoil, and therefore do not look at
alternative paths to large BH retention.  One possibility is that if
the BHs with the appropriate spins merge, some clusters could still
retain larger BHs, even if the recoil velocity of the BHs with no spin is much larger than the escape speed.
On the other hand, spin breaks the symmetry of the binary and
can lead to large recoil velocities even for equal--mass binaries \citep{2004ApJ...607L...5F}.

\subsection{Properties of Merging BHs}
\label{distwave}

\begin{figure}
\plotone{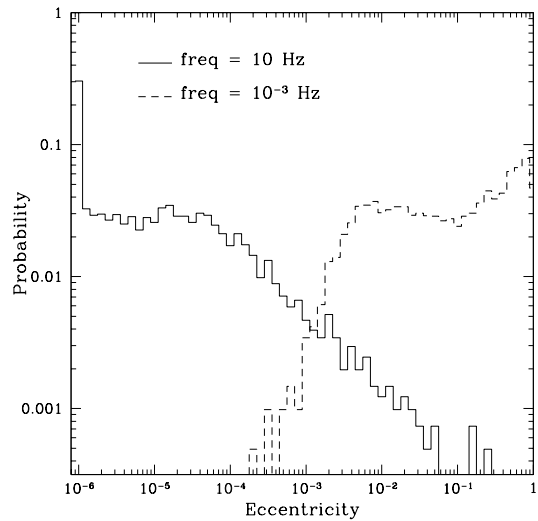}
\caption{Eccentricity distribution of merging BH binaries.  For this
$\log - \log$ plot, we show the eccentricity distribution of all BH
binary mergers for model v2e55k9.  The distribution of eccentricities
is almost entirely independent of the model used. The two frequencies
of the GWs were chosen to show the expected eccentricity distribution
of a binary as it enters the observable bands of both ground--based,
$\approx 10$ Hz, and space--based interferometers, $\sim 10^{-3}$ Hz.  The
low eccentricity of most binaries entering the ground--based
interferometers detection band suggests almost no loss of detectable
BH--BH binary signals if only circular templates are used for
analysis. }
\label{histoecc}
\end{figure}

 Inspiraling BHs with high eccentricity have very complex gravitational
waveforms, and their additional free parameters  make computational
searches even more expensive. Therefore we want to know the
eccentricity of the merging BHs' orbits as they enter possible
detection bands of current and future gravitational wave detectors.
Many ground--based laser interferometers are currently in operation.
These detectors, such as LIGO, are sensitive to GWs
above $\approx 10\,$Hz, below which seismic noise dominates the noise
curve \citep{1994PhRvD..49.2658C}.  Also, in the planning stages is a space--based laser
interferometer, LISA, which, with its longer baseline, is expected to
be sensitive to GWs with frequencies $\sim 10^{-3} - 1\,$Hz \citep{LISAReport,1997CQGra..14.1439B}.

\begin{figure*}
\plottwo{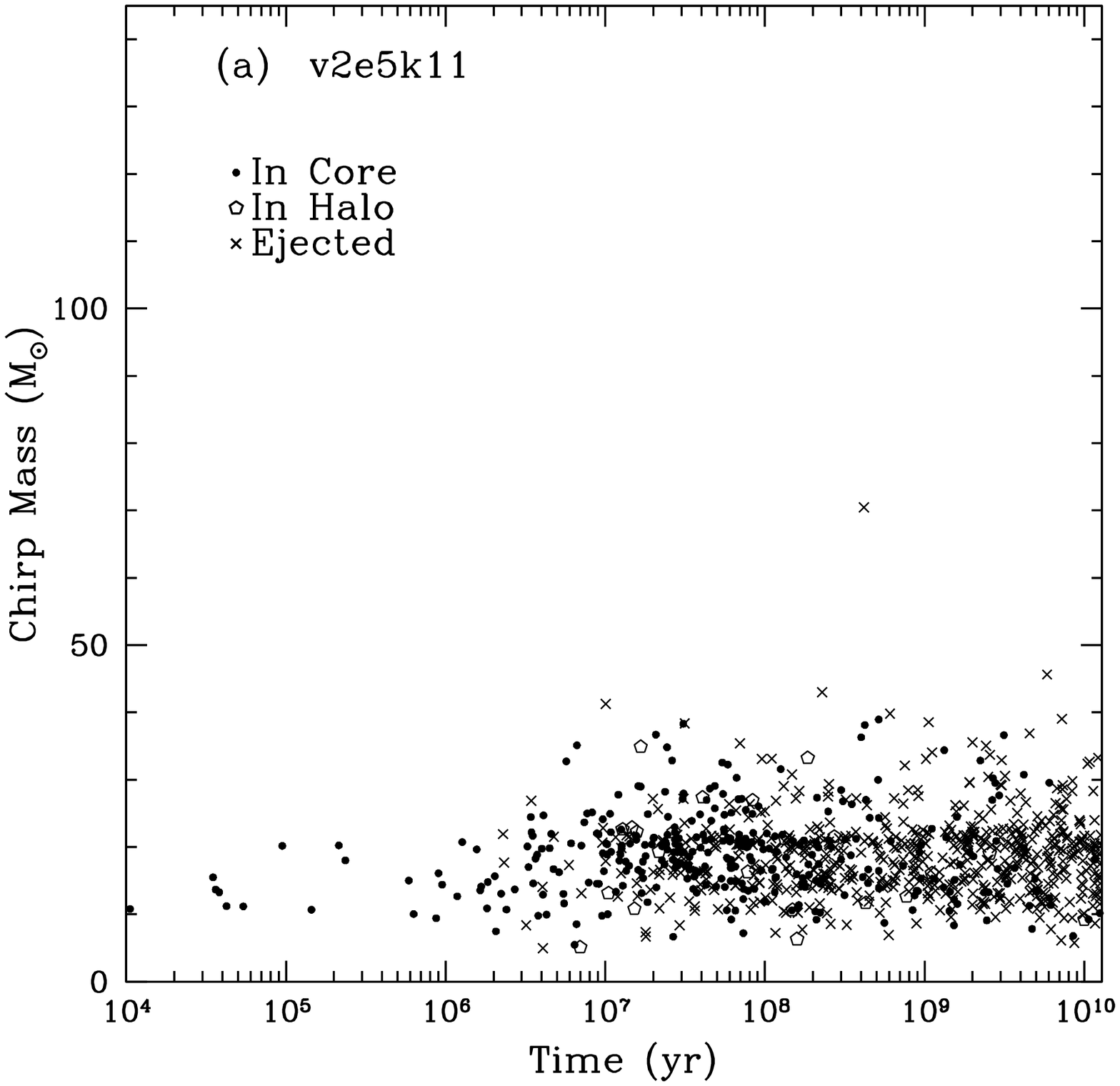}{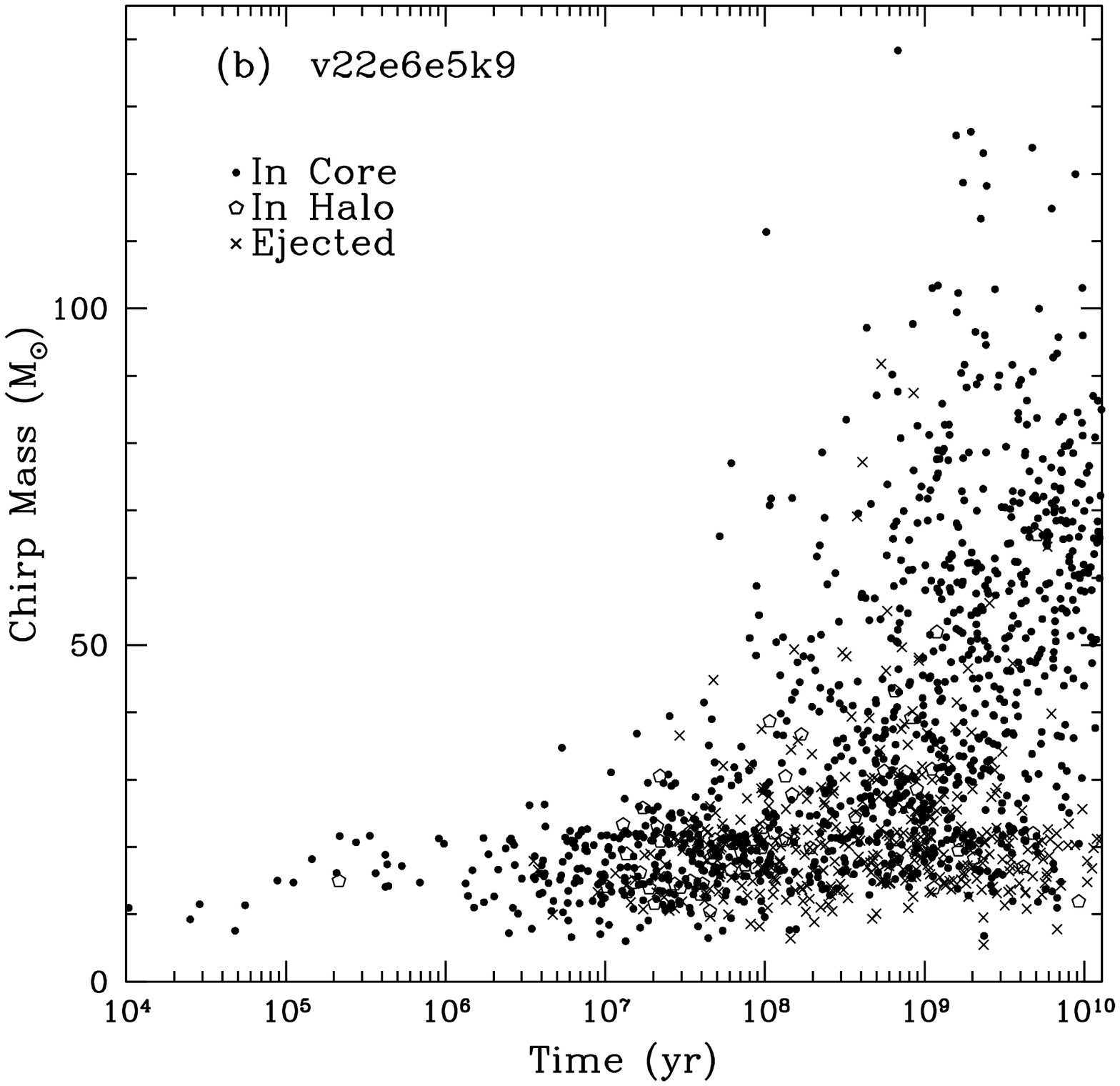}
\caption{Chirp mass of mergers versus time.  This is a comparison of
the two models v2e5k11 and v22e6e5k9, in panels~(a) and (b)
respectively.  Plotted is the chirp mass versus time of
all mergers of 46 random runs of model v2e5k11 and all 46 runs of
v22e6e5k9.  Model v2e5k11 is one of the least efficient clusters in
producing large BHs and BH--BH binary mergers in general. Therefore,
the distribution is most nearly that expected from the initial mass
distribution of BSR04.  Because of how quickly v2e5k11 evolves
($t_{\rm eq} \approx 200\,$Myr) almost all mergers in later times
occur outside the cluster.  In comparison, v22e6e5k9 is a massive
cluster that does not reach equipartition before a Hubble time.
There is still a significant fraction of BHs in the cluster at the end
of the simulation, which allows for more growth, and also more massive
BH mergers.  
}
\label{chirpfig}
\end{figure*}

Dynamical interactions, such as those in a cluster core, coupled with
the strong dependence of merging time with eccentricity, suggest that
many binaries will have highly eccentric orbits after their last
strong interaction.  Of course, GW emission reduces the eccentricity of the orbit
by circularizing the binaries (see eq.~[\ref{pet2}]), and therefore
it is often assumed that most binaries can be fitted with GW templates
for zero eccentricity.  Figure~\ref{histoecc} shows that
there will be almost no loss in possible LIGO BH--BH binary sources
with this assumption because, at such a high frequency, almost all
binaries are circular. However, the eccentricity distribution will
likely matter for space--based detectors since the inspiraling
binaries have not been entirely circularized by the GW emission.  This
is consistent with the previous study by GMH04.

Another important factor in detecting inspirals is the chirp mass of a binary, 
\begin{equation}
M_{\rm chirp} = \frac{(m_1 m_2)^{3/5}}{(m_1+m_2)^{1/5}},
\end{equation}
which solely determines the overall magnitude of the GWs emitted by a
coalescing circular binary. Because our simulations allow for
successive mergers of BHs, as shown in Figure~\ref{chirpfig}, some
mergers have chirp masses well above those expected if dynamics were
not included.  Because of our realistic initial distribution of BH
masses, the chirp masses of most mergers is above the expected value
for two $10\,M_{\odot}$ BHs merging, $8.7\,M_{\odot}$.

Although, as we discussed in \S~\ref{formimbh}, successive mergers of
BH--BH binaries can lead to more inspirals of massive BHs, in most
cases the chirp mass distribution at the end of the evolution of the
cluster is not significantly different from early in the evolution.
For one, about half of the mergers occur outside of the cluster, which
were ejected through dynamic interactions early in the cluster's
history before significant growth had occurred. Also, only in a few
cluster models is the probability of BH growth near unity, in
which the massive BHs dominate the mergers in the cluster.

\begin{figure}
\plotone{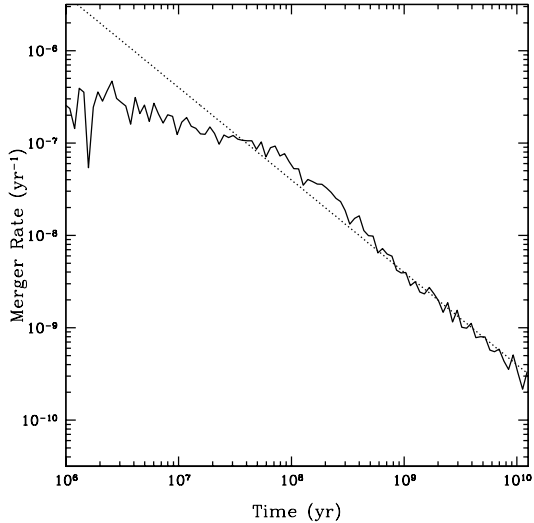}
\caption{Merger rate vs. time.  The solid curve is the average merger
  rate of model v2e55k9 as a function of time.  The dotted line is a  power--law  $\propto\,$time$^{-1}$.  After $\sim 10^8\,$yr, the
  merger rate is inversely proportional to the age of the cluster. The
  evolution of the merger rates can be split into two phases.  The first
  when the cluster is undergoing many binary interactions, and the
  second, when the binary fraction is depleted and nearly zero.
  These two phases of merger rates appear consistently in all cluster models.}
\label{timedist}
\end{figure}

In Figure~\ref{timedist}, we plot the average merger rate of model
v2e55k9 as a function of time.  The merger rate can clearly be broken
into two regimes. The first occurs early in the cluster
evolution, when the number of binaries has not been completely
depleted. The second regime occurs after the cluster is nearly all
single BHs with only a few binaries.  During this later time the
merger rate of the cluster falls off inversely proportional to the age
of the cluster, independent of cluster model.

\subsection{Comparison with Previous Studies}
\label{reswork}

\begin{figure}
\plotone{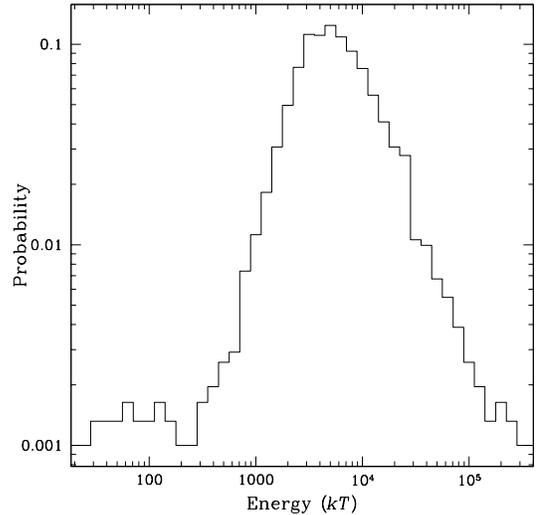}
\caption{Energy distribution of ejected BH binaries.  Plotted is the
probability distribution of the energy of all BH--BH binaries ejected
before equipartition in 117 runs of model v2e55k9.  The energy is plotted in
units of the mean kinetic energy $kT$, where $3/2\,kT$ is the mean
stellar kinetic energy of the MS stars in the core of a cluster of this type. We
find that all other models have a distribution very similar to the one
shown above.}
\label{logeng}
\end{figure}

\begin{figure*}
\plottwo{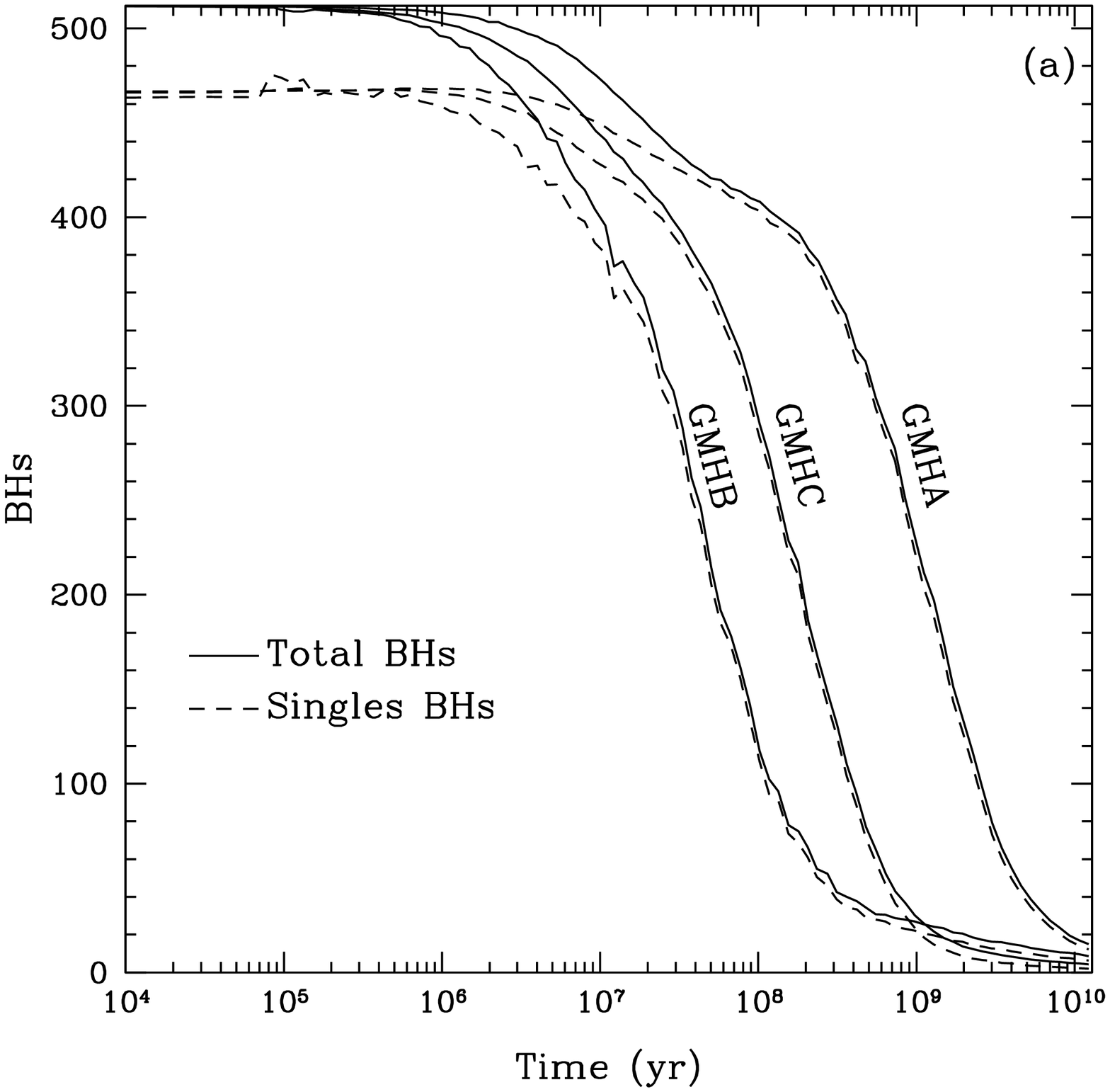}{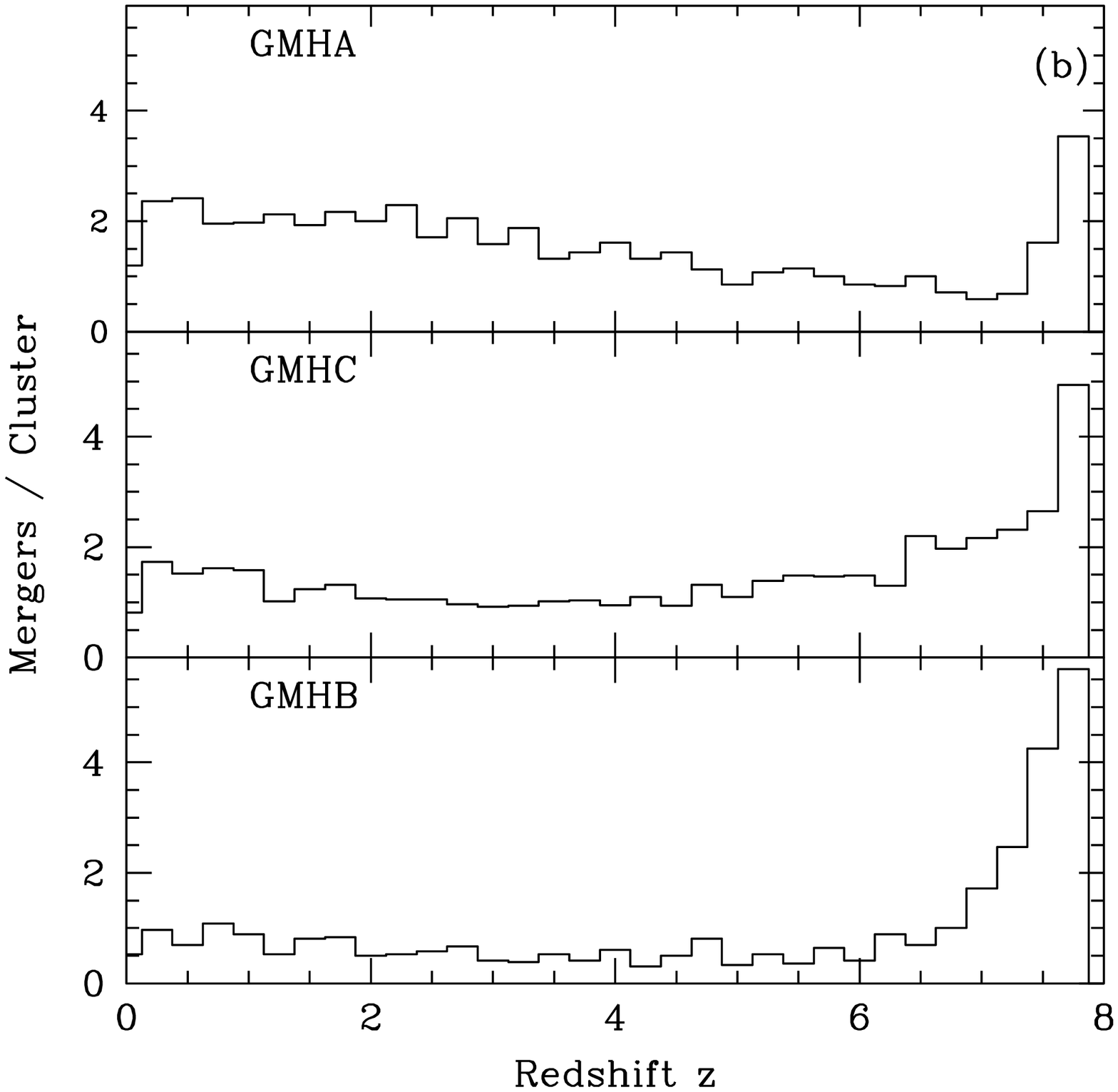}
\caption{A comparison of the models GMHA, GMHB, and GMHC. In panel~(a), the number of BHs as a function of time
is plotted.  Each pair of lines is labeled above by its associated
model that we use to compare our results with GMH04.  Model GMHA, with
only $10\,M_{\odot}$ equal--mass BHs, on average reaches equipartition at $ t_{\rm eq} \approx 5.2\,$Gyr. Model GMHC, a cluster with only $15\,M_{\odot}$ equal--mass
BHs, reaches equipartition at $t_{\rm eq} \approx 830\,$Myr.  Finally, GMHB, with a varied
mass spectrum as in BSR04 and a corresponding average mass $\approx 15\,M_{\odot}$, has not
only a significantly smaller number of ejected binaries, but also
reaches equipartition at the earliest time, $t_{\rm eq} \approx 670\,$Myr.  In panel~(b),
the number of mergers at a given time, denoted by redshift, is
plotted. We assume that the clusters would currently be $13\,$Gyr old.
The model with the varied mass, GMHB, not only dissociates more
quickly, but has most of its mergers at an earlier time in the
cluster's evolution, i.e., at a higher redshift.
\label{mill}}
\end{figure*}

Overall our results agree with those of previous direct $N$-body
simulations.  In the simulations of \citet{2000ApJ...528L..17P}, where
$N = 2048$, $4096$ and $N_{\rm{BH}} = 20$, $40$, approximately $60\%$
of the BHs where ejected as single BHs and $30\%$ ejected as binary
BHs.  The lower values of binary ejection, which we found, can be
explained by in-core mergers and the lower order interactions followed
compared to $N$-body simulations (the calculations of the authors were
Newtonian only, and so did not allow for mergers via GW emission).  As
shown in Table~\ref{tabresults}, in most runs  $70-85\%$ of the
BHs were ejected as singles and $10-15\%$ were ejected in binaries.

The biggest divergence between our simulations and those conducted by
\citet{2000ApJ...528L..17P} is the energy
distribution of ejected BH binaries.  \citet{2000ApJ...528L..17P} found that the
ejected BH binaries had a binding energy distribution more or less uniform in
the logarithm.  In Figure~\ref{logeng} we show a typical binding energy
distribution of the ejected BHs. Our simulations produce a
distribution much more log--normal.  Every model we analyze has a
similar distribution of ejected BH binary binding energy, with values between $\approx 10^3 -
10^5\,kT$, consistent with analytic considerations
\citep{1993Natur.364..421K}.  This divergence can possibly be explained by the low
number of BHs used in \citet{2000ApJ...528L..17P}.

To see what effect the mass spectrum has on our simulations, and also
to compare our data to the results of GMH04, we use three different
models.  Each simulation uses the same velocity dispersion and escape
velocities as in GMH04 (see model GMH in Table~\ref{tab1}), but starts
with a different mass function.  GMHA and GMHC both have
$10\,M_{\odot}$ and $15\,M_{\odot}$ equal--mass BHs respectively.
GMHB has BHs with a mass distribution consistent with our other
simulations following BSR04. This distribution has a corresponding average mass of about
$15\,M_{\odot}$ (see \S \ref{methods} \& \S \ref{init}). As can be
seen in Table~\ref{tabresults}, using equal--mass BHs in our simulations  increases the number of binaries
ejected by almost a factor of two.  The models with equal--mass BHs,
GMHA and GMHC, have about $20\%$ of their BHs ejected as binaries,
whereas GMHB has only about $10\%$.  Figure~\ref{mill} shows how the mass spectrum we use causes the cluster to reach equipartition at
an earlier time and changes the timescales of when mergers occur, compared with a cluster of equal--mass BHs.

\section{Binary BHs as Sources of Gravitational Waves}
\label{gravwave}
BH--BH binaries formed through dynamical interactions may be some of the best sources of GWs detectable by
ground--based laser interferometers.  Previous studies of detection
rates have led to a large range of possible values, with some of the
greatest uncertainty coming from the dynamics of interacting BH
binaries in massive clusters
\citep{1993MNRAS.260..675T,2000ApJ...528L..17P}. In this section we
determine possible maximum detection rates of BH--BH binary inspirals
from some globular cluster models.

\subsection{Calculation of the LIGO Detection Rate}
\label{wavecalc}

To calculate accurately the net detection rate, one should convolve
self--consistent densities and birth rates of observed star clusters
throughout the universe with the mergers rates in our simulations.
Because there is great uncertainty in these distributions, we look at
each of our cluster models, and determine how cluster densities and
masses may affect the distribution of BH--BH inspirals, leaving a
detailed analysis for further study.

Although current ground--based interferometers are not sensitive
enough to detect mergers of inspiraling BH--BH mergers to any
cosmologically significant distance, as these detectors become more sensitive
they will be able to make detections to ever farther distances.  It
then becomes important that a consistent cosmological model is used to
have a better understanding of detectable inspirals.  
In all of our calculations we use the best--fitting cosmological
parameters found by \citet{2003ApJS..148..175S} from combining \emph{WMAP} data
with other measurements: $H_0 = 71\, $km s$^{-1}$ Mpc$^{-1}$,
$\Omega_m = 0.27$, $\Omega_{\gamma} = 5 \times 10^{-5}$, and
$\Omega_{\Lambda} = .73$.

In our calculations, we assume that the globular cluster model
was formed uniformly through the universe at a given cosmological time
corresponding to redshift $z_{\rm form}$. We then record each
detectable merger into one of 100 bins each with time width
$\delta t = t_0/100$, where $t_0$ is the current age of the universe,
based on when the merger occurred.  If $d_i$ is the number of
detections in bin $i$, we sum over the rate of each bin giving the
final rate:
\begin{equation}
R_{z_{\rm form}} = \sum_{i=1}^{100} \frac{d_i}{\delta t} \frac{4\pi}{3} \rho_0 (D_i^3
- D_{i-1}^3) (1+z_i)^{-1},
\label{erate}
\end{equation}
where  $\rho_0$
is the current density of a given cluster model and $z_i$ is the
redshift to bin $i$.  With $t_e = t_0 - i\delta t$, the proper distance to the edge of bin $i$ is 
\begin{equation}
D_i = \int_{t_{\rm e}}^{t_0} \frac{dt}{a(t)},
\label{propdist}
\end{equation}
where $a(t)$ is the scale factor of the Friedman-Robertson-Walker
metric that satisfies the Einstein equation, and $a(t_0) \equiv 1$. The factor $(1+z_i)^{-1}$
in equation~(\ref{erate}) comes from the cosmological time dilation of
the merger rate.

The final detection rate is directly proportional to the density of
such clusters in the universe.  For our calculations we assume that
$\rho_0 = 1\,$Mpc$^{-3}$, independent of cluster model, for ease of
comparing our results with other works. To put this in perspective,
\citet{2000ApJ...528L..17P} found the number density of all globular
clusters to be $\rho_0 \approx 8.4\, h^3\, \rm{Mpc}^{-3}$ based on rough
fits to observations. Our analysis doesn't include the full distribution of
cluster parameters, but instead looks at each cluster individually.  
The Milky Way, for example, contains hundreds of globular clusters,
with only a fraction of clusters similar to those we look at in this
study \citep{1996AJ....112.1487H}.

In order to be as precise as possible, we must determine which mergers
could actually be detected by a given version of LIGO. To do this, we
must look at the accumulated signal--to--noise ratio (SNR) of a given
merger at redshift $z_{\rm m}$. Since, gravitational waveforms are
invariant under redshift, we use the redshifted chirp mass of a given
merger
\begin{equation}
\mathcal{M_{\rm chirp}} =(1+z_{\rm m}) M_{\rm chirp},
\end{equation}
 and its luminosity distance 
\begin{equation}
D_{\rm L} =
(1+z_{\rm m}) D,
\end{equation}
to calculate the SNR of the merger, where $D$ is the proper distance
of the merger as calculated in equation~(\ref{propdist}).

 We look at all BH--BH binary mergers caused by interactions before
the cluster reaches equipartition, and determine if it would be
detected by a given ground--based GW interferometer by comparing its
SNR to that of an inspiraling neutron star--neutron star (NS--NS)
binary at luminosity distance $D_{\rm L,0}$. In practice, we determine
the merger to be detectable if
\begin{equation}
\frac{(S/N)(f_{\rm off})}{(S/N)(D_{\rm L,0})} = \frac{D_{\rm
    L,0}}{D_{\rm L}} \left(\frac{\mathcal{M_{\rm
      chirp}}}{\mathcal{M_{\rm chirp,0}}}\right)^{5/6}
\sqrt{\frac{s(f_{\rm off})}{s(f_{\rm off,0})}} > 1,
\label{snr}
\end{equation}
with  
\begin{equation}
s(f_{\rm off}) = \int_0^{f_{\rm off}} \frac{(f')^{-7/3}}{S_{\rm n}(f')}df',
\end{equation}
where $S_{\rm n}(f')$ is the detector's noise spectrum,
$\mathcal{M_{\rm chirp,0}}$ is the effective chirp mass of the
inspiraling NS--NS binary, and $f_{\rm off,0}$ is the cut--off
frequency of the NS--NS merger \citep{1994PhRvD..49.2658C}. In this
study, we assume that the mergers are isotropic and neglect the
orientation of the merger relative to the detector. We use an analytic
fit for the shape of Advanced LIGO's noise spectrum found in
\citet{1994PhRvD..49.2658C},
\begin{equation}
  S_{\rm n}(f') \propto  \cases{    \infty & $f' < 10\,{\rm Hz},$\cr
    (f_0/f')^4 + 2 [1 + (f'/f_0)^2]  & $f' \geq 10\,{\rm Hz},$}
\label{noise}
\end{equation}
where $f_0 = 70 $ Hz. We do not include the coefficient as calculated by
\citet{1994PhRvD..49.2658C} in this equation 
since it is completely canceled out in equation (\ref{snr}) and
rescaled by $D_{\rm L,0}$.

The final piece of equation (\ref{snr}) is the cut--off frequency of the
merger, $f_{\rm off}$, after which the waveform of the GWs can not be
accurately modeled.  This frequency is generally regarded to be the frequency of GWs at the binary's
last circular orbit (LCO), after which the BHs plunge into each other
in a time less than the orbital period. In their calculations,
\citet*{1993PhRvD..47.3281K} found that for two $10\,M_{\odot}$
Schwarzschild BHs the frequency of the orbit at the LCO is $\approx
100\,$Hz, using what they called ``hybrid'' equations that were valid
through (post)$^{5/2}$--Newtonian order for arbitrary masses.  Because
their calculations also showed that this is a lower limit of the
orbital frequency of the binary for an arbitrary mass ratio, we choose
to use this limit for calculating $f_{\rm off}$.  For circular orbits,
the GW frequency is twice the orbital frequency, therefore in our
calculations we use
\begin{equation}
f_{\rm{off}} \approx 200 \left(\frac{20
  M_{\odot}}{M}\right)\left(\frac{1}{1+z_{\rm m}}\right)\,{\rm Hz}
\end{equation}
as the cut--off frequency of detectable GWs, where $M = m_1+m_2$ and
$z_{\rm m}$ is the redshift of the merger. The location of the LCO and
its corresponding orbital frequency are far from well established.
Nevertheless, numerical simulations and other analytic approximations
have shown the orbital frequencies of equal--mass BHs to be only
larger than the value we use here \citep*{2002PhRvD..65l4009B,
2002PhRvD..65d4021G}.

\subsection{Results}
\begin{figure}
\plotone{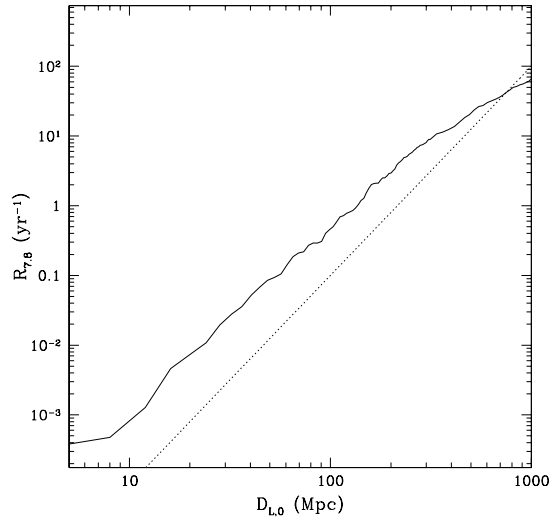}
\caption{Detection rate of BH--BH inspirals.  The solid curve
represents the expected detection rate of mergers from cluster v2e55k9
if it has a current density $\rho_0 = 1\ $Mpc$^{-3}$ and formed at
$z_{\rm i} = 7.84$, or $13\,$Gyr ago.  The method and assumptions of
our calculation are detailed in \S~\ref{gravwave}.  The dotted line is
a power--law for $R \propto D_{\rm L,0}^{-3}$.}
\label{rate}
\end{figure}

We see, in Figure~\ref{rate}, the expected detection rates of BH
binary mergers for cluster v2e55k9.  Here we assume that all clusters
of this model formed when $z_{\rm form} = 7.84$, or $13\,$Gyr ago,
using the assumptions and equations of \S~\ref{wavecalc}.  As can be
seen in the figure, for the luminosity distances Advanced LIGO is
expected to reach, the detection rate scales well by the power--law
$D_{\rm L}^{-3}$. This can be explained by the time evolution of the
merger rate of BHs.  We find, for all simulations, that the merger
rate scales as $t^{-1}$ after the disruption of the primordial
binaries.  For the distances of mergers Advanced LIGO is expected to
be able to detect, the clusters are relatively old and hence, the rate
changes very little.

 The detection rates and theoretical uncertainty of all models can be
found in the last two columns of Table~\ref{tabresults}.  We find for
model v2e55k9, for a version of LIGO capable of detecting NS-NS
mergers at a luminosity distance, $D_{\rm L,0} \approx 190\,{\rm
Mpc}$, the net detection rate is $\approx 2.7\,{\rm yr}^{-1}$.  For
v2e55k9e8, which has the same conditions as v2e55k9, but a recoil
velocity $V_0 = 80\ {\rm km s}^{-1}$, the net detection rate is
actually higher than v2e55k9 at $\approx 4.1\,{\rm yr}^{-1}$. One
cause for this may seem to be the lower cut--off frequency of high mass
binary mergers, but this is wrong.  When analyzing the detection rate
assuming a universal chirp mass for all mergers, the rate was still higher.  It
can actually be attributed to the higher merger rate later in the
evolution of the cluster.  This delayed merger rate may be a result of
a few mechanisms.  The dependence of merger rate in models v2e55k9-100
and v2e55k9-200 suggests that having more massive objects in the
cluster results in a lower rate overall at the end of the evolution.
Model v2e55k9-200, which has a $200\,M_{\odot}$ seed BH, has a merger
about half that of model v2e55k9.  Another, slightly less significant
mechanisms can still possibly be the masses of the merging
binaries. The timescale for merger is longer when the masses of the objects are less (see eqs. [\ref{pet1}] and [\ref{pet2}]).

Within the level of theoretical uncertainty, the final detection rate
is roughly proportional to the initial number of BHs.  This, of
course, is not exact, and isn't expected to be.  The timescale for
cluster disruption is also proportional to the number of BHs, so it
should be expected that there would be some variance in the merger
rate at later times depending on the initial number of BHs.

In order to get an estimated value of the actual Advanced LIGO
detection rate, we must consider not only the number density of the
massive globular clusters we look at here, but also the density of the
low mass clusters.  Our results suggest that the expected detection
rate scales proportionally to the initial number of BHs in the
cluster.  Considering the number of BHs that
\citet{2000ApJ...528L..17P} quoted as being in the globular clusters
they analyzed, we would expect the actual rate to lie within the range
of rates for the clusters reported in Table~\ref{tabresults}, rescaled to the
number density of globular clusters in the universe ($\sim 3\,$Mpc$^{-3}$).

\section{Summary and Discussion}
\label{discussion}

Our work is based on the reasonable assumption that, in a sufficiently
large and dense star cluster, 
BHs created via stellar evolution concentrate in an inner core and
effectively decouple from the rest of the cluster following mass
segregation and the development of the Spitzer instability. Taking
advantage of this decoupling, we have computed the dynamical evolution
of the BHs in a highly simplified treatment of the stellar dynamics
but covering a wide range of cluster models and repeating calculations
for each model in order to obtain a complete statistical description
of outcomes. Our assumed initial distributions of BH masses and binary
parameters are based on the most recent population synthesis
calculations for young stellar systems. In our simulations, we use a
simple Monte Carlo method to follow the evolution of the BH subsystem
in a fixed background cluster described by a King model. The BHs are
allowed to interact only in the core, described by a constant density
and velocity dispersion. All interactions involving binaries are
computed exactly by direct (\Fewbody) integrations but we implement
three-body binary formation using a simple analytic rate formula
(eq.~[\ref{threebody}]).  Binary formation through dissipative two-body
encounters is negligible in the systems we consider here (with
velocity dispersions $\sigma < 100\, {\rm km}\,{\rm s}^{-1}$). We
allow for Newtonian recoil of BHs into the halo, reintroducing them in
the core following mass segregation.  Between interactions, BH--BH
binaries are evolved taking into account gravitational radiation and
the possibility of a merger (eqs.~[\ref{pet1}] \& [\ref{pet2}]). In some
simulations, we attempt to model kicks due to gravitational radiation
recoil in merging binaries, parameterizing the large theoretical
uncertainty in a simple analytic formula that depends on the binary
mass ratio only (eq.~[\ref{recoil}]).

We present the results of our simulations in \S~\ref{results} and
\ref{gravwave}, and, in particular, we derive the probability of
massive BH growth and retention in clusters ($\sim 20 - 80\%$).  We
also show that the Kozai mechanism in triples has almost no
significant effect on the merger rate, BH growth, or retention, in
contrast to previous suggestions in the literature. In addition, we
derive a net rate of BH--BH binary mergers detectable by current and
future ground--based GW interferometers. We find, under extremely
optimistic assumptions, that this can be up to a few tens of events
per year, if we assume that the globular clusters likely formed over a
time span $z_{\rm form} \la 8$.  When including recoil from the
gravitational ``rocket'' effect \citep{1983MNRAS.203.1049F} this rate
is nearly doubled due to the increased merger rate from a population
of less massive BHs.

The simulations presented here improve upon previous studies in that
we not only account for binary evolution due to GW emission between
successive interactions of a fixed group of BHs, but we also use a
realistic BH IMF based on the most recent population synthesis models
(BSR04).  Most analytic studies and numerical simulations so far
assumed that all BHs were $10\,M_{\odot}$ (see, e.g.,
\citealt{2002MNRAS.330..232M} and \citealt{2000ApJ...528L..17P}), or
included just one large BH (GMH04).  The mass spectrum from BSR04
gives a slightly higher average mass ($\approx 15\,M_{\odot}$), and
also includes at least one significantly large BH, which is almost
always ejected early in our simulations.  Even some of the largest BHs
formed from mergers, with mass $\approx 120\,M_{\odot}$, are ejected
from the cluster when they are formed early enough in the simulation
to interact often with other large BHs.  The distribution of the
binary parameters also has a significant effect on the interactions.
The low primordial binary fraction results in very few formed triples,
and even fewer mergers in triples, whether enhanced by the Kozai
mechanism or not.

The results of our simulations indicate a greater likelihood of
moderate growth in globular clusters than previous $N$-body
simulations have suggested possible, but they show varied results
compared with GMH04. Our wider BH mass function makes the ejection of
BHs with masses as high as $\sim 100\,M_{\odot}$ very likely.  On the
other hand it also encourages the formation of even larger BHs through successive
mergers, and the chirp masses of merging binaries are larger than
would be expected for a cluster with equal--mass $\sim 10\,M_\odot$
BHs.

Despite the demonstrated growth, the probability for an IMBH of
$\sim~10^3\,M_{\odot}$ to form directly via successive BH--BH mergers
remains extremely small. However, we should not neglect the
possibility of significant further growth of the final remaining BH
through stellar collisions after complete evaporation of the BH
sub-cluster.  Results from \citet*{2004ApJ...613.1133B} suggest that
even a moderately massive $\sim 200\,M_{\odot}$ BH could grow into a
$10^3\,M_{\odot}$ IMBH after only a few Gyr, well within the current
ages of globular clusters.

Initial conditions play a crucial role in determining the probability
of IMBH growth.
In general, massive, dense clusters with high core temperatures have
 the greatest likelihood of BH growth.  Eccentricity growth through
 the Kozai mechanism, although it makes it possible for BHs to merge
 with very little (Newtonian) recoil, does not occur often enough to
 affect IMBH formation.  Note, however, that our basic assumptions
 break down at late times, when the growth of BHs is most
 significant. Eventually the number of remaining BHs becomes small
 enough that the constant core conditions assumed in our simulations
 are no longer justified. In particular, the BH subsystem will start
 interacting with lower-mass stars at a significant rate and it will
 then likely return to energy equipartition. This is why we emphasize
 the values computed by our code at equipartition, as opposed to the
 values at the end of the evolution (\S~\ref{bhform}). Beyond the
 return to equipartition, the evolution of the remaining BHs will be
 strongly coupled to the overall evolution of the whole star cluster.

Our simulations are far from complete, yet the results show great promise for future work.
For one, we can expect that the most massive clusters will have the
greatest likelihood of  BH growth.  To enter the next regime of even
more massive clusters, such as galactic nuclei, one must
account for binary formation from  gravitational bremsstrahlung
\citep{1993ApJ...418..147L}. Also, because of the simplifications we make,
it is also possible to explore an even larger parameter space,
especially of the more numerous and less massive clusters.  By convolving
these data with the cluster formation history of the universe, one
could determine a BH--BH merger detection rate to an ever more
accurate degree. 

\acknowledgments We would like to thank Marc Freitag for providing us
with the code to determine the King model parameters used in our
simulations. This work was supported by NASA Grant NAG5-12044 and NSF
Grant NSF PHY-0245028. F.A.R. thanks the Center for Gravitational Wave
Physics at Penn State for hospitality and support.

\clearpage
\begin{landscape}
\begin{deluxetable*}{lcccccccccccccc}
\tabletypesize{\scriptsize}
\tablecolumns{15}
\tablewidth{0pc}
\tablecaption{Results}
\tablehead{
\colhead{}&
\colhead{}&
\colhead{Avg. Largest}&
\colhead{Std. Dev.}&
\colhead{Number of}&
\colhead{Largest BH}&
\colhead{Fraction}&
\colhead{}&
\colhead{}&
\colhead{$\log_{10}t_{\rm eq}$}&
\colhead{}&
\colhead{}&
\colhead{}&
\colhead{$R_{7.8} \pm \sigma$}&
\colhead{$R_1 \pm \sigma$}
\\
\colhead{Model}&
\colhead{Runs}&
\colhead{BH Mass ($M_{\odot}$)}&
\colhead{($M_{\odot}$)}&
\colhead{BH Mergers}&
\colhead{Mass ($M_{\odot}$)}&
\colhead{$> 100\,M_{\odot}$}&
\colhead{$f_{\rm e,sin}$}&
\colhead{$f_{\rm e,bin}$}&
\colhead{(yr)}&
\colhead{$\frac{N_{\rm trip}}{N_{\rm cluster}}$}&
\colhead{$\frac{N_{\rm cluster}}{N_{\rm merge}}$}&
\colhead{$N_{\rm merge}$}&
\colhead{(yr$^{-1}$)}&
\colhead{(yr$^{-1}$)}
}
\startdata
e5e5king7\tablenotemark{a}\dotfill   &  $\phn$99   &  $\phn$62   &   $\phn$20    &   1.0   &   164   &   0.05   &   0.15   &   0.03   &  $\phn$\nodata\tablenotemark{a}   &   0.01   &   0.51   &   10 & 0.78 $\pm$ .17  & 1.6 $\pm$ .3 \\ 
v2e5k7\tablenotemark{a}\dotfill   &   $\phn$99   &   147   &  $\phn$99&  5.5   &   370   &   0.59   &   0.73   &   0.08   &   $\phn$\nodata\tablenotemark{a}   &   0.02   &   0.53   &   30 &     3.2 $\pm$ .4& 10.3 $\pm$ .7\\ 
v3e5k7\dotfill   &   $\phn$65   &   144   &  $\phn$98 &  5.5   &   395   &   0.60  &   0.80  &   0.10   &   9.69   &   0.04   &   0.51   &   35 &  2.1 $\pm$ .3 & 5.5 $\pm$ .6 \\ 
v3e5k7ej54\dotfill   &   $\phn$53   &   123   &  100 &  4.5   &   370   &   0.47   &   0.79   &   0.10   &   9.58   &   0.04   &   0.51   &   37 & 2.3 $\pm$ .4 & 5.7 $\pm$ .7 \\ 
v3e5k7ej75\dotfill   &   $\phn$62   &   $\phn$32   &  $\phn$15  & 0.6   &   107   &   0.02   &   0.77   &   0.12   &   9.46   &   0.04   &   0.48  &   41 &  4.0 $\pm$ .5  &  6.8 $\pm$ .7\\ 
v5e5k7\dotfill   &   $\phn$26   &   $\phn$81   &  $\phn$66&  3.2   &   303   &   0.38   &   0.77   &   0.12   &   8.63   &   0.13   &   0.49   &   44 &  1.7 $\pm$ .5 &  5.7 $\pm$ .9\\ 
e5e5king9\tablenotemark{a}\dotfill   &   $\phn$99   &   $\phn$66   &  $\phn$34  & 1.4   &   246   &   0.14   &   0.33   &   0.05   &   $\phn$\nodata\tablenotemark{a}   &   0.01   &   0.43   &   13 & 2.6 $\pm$ .3  & 3.7 $\pm$ .4 \\ 
v2e5k9\dotfill   &   $\phn$96   &   $\phn$71   &  $\phn$63  &  2.3   &   269   &   0.22   &   0.78   &   0.12   &   9.66   &   0.04   &   0.41   &   28 & 3.6 $\pm$ .5  &  6.6 $\pm$ .5\\ 
v3e5k9\dotfill   &   $\phn$27   &   $\phn$52   &  $\phn$48 &  1.4   &   209   &   0.19   &   0.77   &   0.13   &   8.86   &   0.11   &   0.41   &   30 &  4.2 $\pm$ .9 & 5.1 $\pm$ .9 \\ 
e55king9\tablenotemark{a}\dotfill   &   100   &   106   &   $\phn$73 &  3.4   &   330   &   0.40  &   0.49   &   0.05   &   $\phn$\nodata\tablenotemark{a}   &   0.01   &   0.49   &   19 & 4.8 $\pm$ .5  & 7.2 $\pm$ .5\\ 
v2e55k9--256\dotfill   &   $\phn$20   &   $\phn$61   &  $\phn$54 & 1.7  &   219   &   0.15   &   0.73   &   0.10   &   9.08   &   0.07   &   0.47   &   18 & 0.96 $\pm$ .39  &  1.3 $\pm$ .5\\ 
v2e55k9\dotfill   &   117   &   104   &  $\phn$80 &  3.8   &   296   &   0.43   &   0.80  &   0.10   &   9.39   &   0.03   &   0.43   &   35&  2.7 $\pm$ .3 & 5.7 $\pm$ .5 \\ 
v2e55k9--1024\dotfill   &   $\phn$38   &   160   & 134 &  6.5   &   462   &   0.55   &   0.83   &   0.10   &   9.59   &   0.02   &   0.41   &   69 &  4.4 $\pm$ .7 & 10 $\pm$ 1 \\ 
v2e55k9--2048\dotfill   &   $\phn$37   &   208   &   164 & 9.2   &   619   &   0.68   &   0.85   &   0.10   &   9.72   &   0.03   &   0.40  &   133$\phn$& 9.9 $\pm$ 1.1  & 19 $\pm$ 1 \\ 
v2e55k9--100\dotfill   &   $\phn$57   &   143   &  123  &  4.9   &   405   &   0.49   &   0.81   &   0.09   &   9.38   &   0.02   &   0.47   &   33 & 1.7 $\pm$ .4  & 4.8 $\pm$ .6 \\ 
v2e55k9--200\dotfill   &   $\phn$43   &   212   &  186 &  5.6   &   519   &   0.56   &   0.82   &   0.08   &   9.34   &   0.02   &   0.49   &   29 & 1.0 $\pm$ .7  &  3.8 $\pm$ .6\\ 
v2e55k9e6\dotfill   &   $\phn$40   &   $\phn$80   &  $\phn$87 &  2.6   &   308   &   0.23   &   0.79   &   0.11   &   9.38   &   0.03   &   0.44   &   37 &  2.0 $\pm$ .5 & 6.6 $\pm$ .9 \\ 
v2e55k9e65\dotfill   &   $\phn$40   &   $\phn$56   &  $\phn$69 & 1.6  &   285   &   0.13   &   0.78   &   0.12   &   9.32   &   0.03   &   0.41   &   39 &  2.9 $\pm$ .6 & 7.3 $\pm$ .9 \\ 
v2e55k9e7\dotfill   &   $\phn$39   &   $\phn$37   &   $\phn$37 & 0.7   &   217   &   0.05   &   0.78   &   0.12   &   9.31   &   0.05   &   0.40  &   40 &  3.4 $\pm$ .6 &  8 $\pm$ 1\\ 
v2e55k9e8\dotfill   &   $\phn$40   &   $\phn$37   &   $\phn$30 & 0.7   &   186   &   0.05   &   0.77   &   0.12   &   9.26   &   0.04   &   0.40  &   40 &  4.1 $\pm$ .6 &  7.9 $\pm$ .9\\ 
v3e55k9\dotfill   &   $\phn$19   &   $\phn$66   &  $\phn$70 &  2.2   &   276   &   0.21   &   0.77   &   0.13   &   8.47   &   0.11   &   0.36   &   40 &  4.5 $\pm$ 1.0 & 6.2 $\pm$ 1.0 \\ 
v2e6e5k9\dotfill   &   $\phn$46   &   248   & 105 &  10.3$\phn$  &   445   &   0.89   &   0.81   &   0.06   &   10.04$\phn$  &   0.02   &   0.59   &   35 & 0.94 $\pm$ .28  &  3.9 $\pm$ .6\\ 
v3e6e5k9\dotfill   &   $\phn$37   &   199  &  115 &  7.9   &   386   &   0.76   &   0.81   &   0.08   &   9.24   &   0.03   &   0.53   &   40 & 1.0 $\pm$ .3  &  3.5 $\pm$ .7\\ 
v22e6e5k9\tablenotemark{a}\dotfill   &   $\phn$46   &   367   &  102 &  14.8$\phn$  &   547   &   0.98   &   0.72   &   0.03   &   $\phn$\nodata\tablenotemark{a}   &   0.02   &   0.76   &   33 & 0.57 $\pm$ .24 & 2.9 $\pm$ .5 \\ 
v32e6e5k9\dotfill   &   $\phn$47   &   373   &  $\phn$95 &  15.6$\phn$  &   509   &   1.00   &   0.83   &   0.05   &   9.81   &   0.02   &   0.72   &   39& 0.66 $\pm$ .25  &  1.2 $\pm$ .3\\ 
e5king9\tablenotemark{a}\dotfill   &   $\phn$99   &   145   &   $\phn$84  &  5.0   &   335   &   0.61   &   0.56   &   0.05   &  $\phn$\nodata\tablenotemark{a}   &   0.02   &   0.51   &   21 & 4.2 $\pm$ .4  & 7.5 $\pm$ .6 \\ 
v2o5k9\dotfill   &   $\phn$35   &   138   &  $\phn$91 & 5.2   &   353   &   0.60  &   0.81   &   0.09   &   9.29   &   0.03   &   0.46   &   37 & 2.0 $\pm$ .5  & 4.5 $\pm$ .8 \\ 
v3o5k9\dotfill   &   $\phn$21   &   107   &  $\phn$88 &  4.1  &   308   &   0.43   &   0.78   &   0.12   &   8.31   &   0.07   &   0.41   &   42& 2.3 $\pm$ .7  &  6 $\pm$ 1\\ 
e5e5king11\dotfill   &   $\phn$99   &   $\phn$40   &  $\phn$24 &  0.8   &   142   &   0.05   &   0.76   &   0.12   &   10.09$\phn$  &   0.03   &   0.41   &   16 & 3.3 $\pm$ .4  & 6.5 $\pm$ .5 \\ 
v2e5k11\dotfill   &   $\phn$49   &   $\phn$41   &  $\phn$18 &  0.8   &   $\phn$97   &   0.02   &   0.76   &   0.15   &   8.76   &   0.11   &   0.37   &   20 & 2.4 $\pm$ .5  &  5.1 $\pm$ .7\\ 
v3e5k11\dotfill   &   $\phn$14   &   $\phn$42   &  $\phn$15 &  0.7   &   $\phn$75   & 0.00 &   0.75   &   0.16   &   8.42   &   0.24   &   0.41   &   26 & 2.1 $\pm$ .9  &  4 $\pm$ 1 \\ 
e55king11\dotfill   &   $\phn$99   &   $\phn$55   &  $\phn$42 & 1.5   &   237   &   0.13   &   0.79   &   0.12   &   9.88   &   0.02   &   0.38   &   24 & 3.5 $\pm$ .4  & 8.7 $\pm$ .6 \\ 
v2e55k11\dotfill   &   $\phn$30   &  $\phn$40   & $\phn$21  &  0.6  &   117   &   0.07   &   0.77   &   0.15   &   8.43   &   0.09   &   0.31   &   29 & 3.0 $\pm$ .6  & 8 $\pm$ 1 \\ 
v3e55k11\dotfill   &   $\phn$$\phn$8   &   $\phn$52   &  $\phn$19 &  1.4   &   $\phn$89   & 0.00 &   0.75   &   0.15   &   8.08   &   0.17   &   0.40  &   34 & 1.6 $\pm$ .8  & 5 $\pm$ 2 \\ 
e5king11\dotfill   &   $\phn$99   &  $\phn$62   & $\phn$57 &  1.7   &   297   &   0.18   &   0.80  &   0.11   &   9.80  &    0.03   &    0.37   &   27 & 3.3 $\pm$ .4  &  8.6 $\pm$ .6\\ 
v2o5k11\dotfill   &   $\phn$53   &   $\phn$53   & $\phn$36 &  1.3   &   202   &   0.17   &   0.78   &   0.14   &   8.29   &   0.07   &   0.33   &   32 & 2.7 $\pm$ .5  &  5.3 $\pm$ .7\\ 
v2o5k1110\dotfill   &  $\phn$20   &   $\phn$52   &  $\phn$43 & 1.4  &   210   &   0.20  &   0.80  &   0.14   &   8.51   &   0.06   &   0.30  &   66& 8.0 $\pm$ 1.4  &  14 $\pm$ 2\\ 
v2o5k1111\dotfill   &   $\phn$$\phn$9   &   $\phn$75   &  $\phn$91  & 2.6   &   298   &   0.22   &   0.82   &   0.14   &   8.74   &   0.06   &   0.28   & 131$\phn$&  12 $\pm$ 2 &  31 $\pm$ 4\\ 
v3o5k11\dotfill   &   $\phn$$\phn$8   &   $\phn$61   &  $\phn$27 &  1.3   &   100   & 0.00 &   0.74   &   0.17   &   7.97   &   0.15   &   0.31   &   40& 4.4 $\pm$ 1.5  &  10 $\pm$ 3\\ 
v2e5e7k11\dotfill   &  $\phn$30   &   $\phn$88  & $\phn$63 &  3.1  &   239   &   0.43   &   0.79   &   0.12   &   7.81   &   0.06   &   0.36   &   43 & 1.6 $\pm$ .5  &  5.7 $\pm$ .9\\ 
v3e5e7k11\dotfill   &  $\phn$$\phn$8   &  $\phn$41   &  $\phn$10 & 0.8   &   $\phn$56   & 0.00 &   0.74   &   0.16   &   7.46   &   0.14   &   0.34   &   53 & 1.8 $\pm$ .9  & 5 $\pm$ 2 \\ 
GMHA\dotfill   &   $\phn$41   &   $\phn$20   &  $\phn$11   &  1.0   &   $\phn$50   & 0.00 &   0.69   &   0.21   &   9.71   &   0.03  &   0.35   &   49& 2.1 $\pm$ .4  &  5.8 $\pm$ .6\\ 
GMHB\dotfill   &   $\phn$36   &  $\phn$68   &   $\phn$69  &  2.3   &   291   &   0.22   &   0.79   &   0.12   &   8.82   &   0.05   &   0.38   &   32 & 3.4 $\pm$ .7  &  6.1 $\pm$ .9\\ 
GMHC\dotfill   &  $\phn$60   &  $\phn$31   & $\phn$22   & 1.1   &   120   &   0.03   &   0.68   &   0.23   &   8.92   &   0.03   &   0.30  &    47&    3.9 $\pm$ .5 & 8.4 $\pm$ .7\\ 
\enddata
\tablecomments{Results of all simulations. Starting from the left the columns are the model name, number of total runs made, the average mass of the largest remaining BH at equipartition, the standard deviation of the largest mass, the total number of successive mergers the average largest remaining BH went through before  equipartition, the largest BH  formed that remained in the cluster, the fraction of simulations with a BH off mass $> 100\,M_{\odot}$ present at equipartition, the fraction of BHs ejected as singles ($f_{\rm e,sin}$), the fraction of BHs ejected in binaries ($f_{\rm e,bin}$), the log of time at which the cluster reached equipartition($t_{\rm eq}$), the fraction of mergers in the cluster that happened in triples ($N_{\rm trip}/N_{\rm cluster}$), the fraction of total mergers that occurred in the cluster ($N_{\rm cluster}/N_{\rm merge}$), the average total number of mergers for a single simulation ($N_{\rm merge}$), the expected Advanced LIGO rate of detection if the cluster formed when $z_{\rm form} = 7.8$ ($R_{7.8}$), and the expected Advanced LIGO detection rate if the cluster formed when  $z_{\rm form} = 1$ ($R_{1}$).  For our calculation of the expected detectable merger rate, we assume that the given cluster model has a current uniform number density $\rho_0 = 1\,$Mpc$^{-3}$ and that the interferometer is capable of detecting a NS--NS inspiral up to $D_{\rm L,0} = 190\,$Mpc (see \S \ref{gravwave} for our detailed calculations and assumptions).   }
\label{tabresults}
\tablenotetext{a}{In these simulations the cluster never reaches equipartition before a Hubble time.  All numbers are calculated at the end of the simulation at $\log_{10}{t_{\rm{eq}}} = 10.13$.}
\end{deluxetable*}
\clearpage
\end{landscape}


\begin{thebibliography}{61}
\expandafter\ifx\csname natexlab\endcsname\relax\def\natexlab#1{#1}\fi

\bibitem[{{Baumgardt} {et~al.}(2003{\natexlab{a}}){Baumgardt}, {Hut}, {Makino},
  {McMillan}, \& {Portegies Zwart}}]{2003ApJ...582L..21B}
{Baumgardt}, H., {Hut}, P., {Makino}, J., {McMillan}, S., \& {Portegies Zwart},
  S. 2003{\natexlab{a}}, \apjl, 582, L21

\bibitem[{{Baumgardt} {et~al.}(2004{\natexlab{a}}){Baumgardt}, {Makino}, \&
  {Ebisuzaki}}]{2004ApJ...613.1133B}
{Baumgardt}, H., {Makino}, J., \& {Ebisuzaki}, T. 2004{\natexlab{a}}, \apj,
  613, 1133

\bibitem[{{Baumgardt} {et~al.}(2004{\natexlab{b}}){Baumgardt}, {Makino}, \&
  {Ebisuzaki}}]{2004ApJ...613.1143B}
---. 2004{\natexlab{b}}, \apj, 613, 1143

\bibitem[{{Baumgardt} {et~al.}(2005){Baumgardt}, {Makino}, \&
  {Hut}}]{2005ApJ...620..238B}
{Baumgardt}, H., {Makino}, J., \& {Hut}, P. 2005, \apj, 620, 238

\bibitem[{{Baumgardt} {et~al.}(2003{\natexlab{b}}){Baumgardt}, {Makino}, {Hut},
  {McMillan}, \& {Portegies Zwart}}]{2003ApJ...589L..25B}
{Baumgardt}, H., {Makino}, J., {Hut}, P., {McMillan}, S., \& {Portegies Zwart},
  S. 2003{\natexlab{b}}, \apjl, 589, L25

\bibitem[{{Belczynski} {et~al.}(2004){Belczynski}, {Sadowski}, \&
  {Rasio}}]{2004ApJ...611.1068B}
{Belczynski}, K., {Sadowski}, A., \& {Rasio}, F.~A. 2004, \apj, 611, 1068

\bibitem[{{Bender} {et~al.}(1998){Bender}, {Brillet}, {Ciufolini}, {Cruise},
  {Cutler}, {Danzmann}, {Fidecaro}, {Folkner}, {Hough}, {McNamara},
  {Peterseim}, {Robertson}, {Rodrigues}, {R{\" u}diger}, {Sandford}, {Sch{\"
  a}fer}, {Schilling}, {Schutz}, {Speake}, {Stebbins}, {Sumner}, {Touboul},
  {Vinet}, {Vitale}, {Ward}, \& {Winkler}}]{LISAReport}
{Bender}, P., {Brillet}, A., {Ciufolini}, I., {Cruise}, A.~M., {Cutler}, C.,
  {Danzmann}, K., {Fidecaro}, F., {Folkner}, W.~M., {Hough}, J., {McNamara},
  P., {Peterseim}, M., {Robertson}, D., {Rodrigues}, M., {R{\" u}diger}, A.,
  {Sandford}, M., {Sch{\" a}fer}, G., {Schilling}, R., {Schutz}, B., {Speake},
  C., {Stebbins}, R.~T., {Sumner}, T., {Touboul}, P., {Vinet}, J.-Y., {Vitale},
  S., {Ward}, H., \& {Winkler}, W. 1998, LISA Pre-Phase A Report, 2nd ed.

\bibitem[{{Bender} \& {Hils}(1997)}]{1997CQGra..14.1439B}
{Bender}, P.~L., \& {Hils}, D. 1997, Classical and Quantum Gravity, 14, 1439

\bibitem[{{Blanchet}(2002)}]{2002PhRvD..65l4009B}
{Blanchet}, L. 2002, \prd, 65, 124009

\bibitem[{{Blanchet} {et~al.}(2005){Blanchet}, {Qusailah}, \& {Will}}]{2005Blanchetetal}
  {Blanchet}, L., {Qusailah}, M.~S.~S., \& {Will}, C.~M. 2005, \apj, accepted (astro-ph/0507692)
 
\bibitem[{{Bonnell} {et~al.}(2001){Bonnell}, {Bate}, {Clarke}, \&
  {Pringle}}]{2001MNRAS.323..785B}
{Bonnell}, I.~A., {Bate}, M.~R., {Clarke}, C.~J., \& {Pringle}, J.~E. 2001,
  \mnras, 323, 785

\bibitem[{{Colgate}(1967)}]{1967ApJ...150..163C}
{Colgate}, S.~A. 1967, \apj, 150, 163

\bibitem[{{Cutler} \& {Flanagan}(1994)}]{1994PhRvD..49.2658C}
{Cutler}, C., \& {Flanagan}, {\' E}.~E. 1994, \prd, 49, 2658

\bibitem[{{Ebisuzaki} {et~al.}(2001){Ebisuzaki}, {Makino}, {Tsuru}, {Funato},
  {Portegies Zwart}, {Hut}, {McMillan}, {Matsushita}, {Matsumoto}, \&
  {Kawabe}}]{2001ApJ...562L..19E}
{Ebisuzaki}, T., {Makino}, J., {Tsuru}, T.~G., {Funato}, Y., {Portegies Zwart},
  S., {Hut}, P., {McMillan}, S., {Matsushita}, S., {Matsumoto}, H., \&
  {Kawabe}, R. 2001, \apjl, 562, L19

\bibitem[{{Favata} {et~al.}(2004){Favata}, {Hughes}, \&
  {Holz}}]{2004ApJ...607L...5F}
{Favata}, M., {Hughes}, S.~A., \& {Holz}, D.~E. 2004, \apjl, 607, L5

\bibitem[{{Fitchett}(1983)}]{1983MNRAS.203.1049F}
{Fitchett}, M.~J. 1983, \mnras, 203, 1049

\bibitem[{{Ford} {et~al.}(2000){Ford}, {Kozinsky}, \&
  {Rasio}}]{2000ApJ...535..385F}
{Ford}, E.~B., {Kozinsky}, B., \& {Rasio}, F.~A. 2000, \apj, 535, 385; erratum
  605, 966

\bibitem[{{Fregeau} {et~al.}(2004){Fregeau}, {Cheung}, {Portegies Zwart}, \&
  {Rasio}}]{2004MNRAS.352....1F}
{Fregeau}, J.~M., {Cheung}, P., {Portegies Zwart}, S.~F., \& {Rasio}, F.~A.
  2004, \mnras, 352, 1

\bibitem[{{Fregeau} {et~al.}(2003){Fregeau}, {G{\" u}rkan}, {Joshi}, \&
  {Rasio}}]{2003ApJ...593..772F}
{Fregeau}, J.~M., {G{\" u}rkan}, M.~A., {Joshi}, K.~J., \& {Rasio}, F.~A. 2003,
  \apj, 593, 772

\bibitem[{{Fregeau} {et~al.}(2002){Fregeau}, {Joshi}, {Portegies Zwart}, \&
  {Rasio}}]{2002ApJ...570..171F}
{Fregeau}, J.~M., {Joshi}, K.~J., {Portegies Zwart}, S.~F., \& {Rasio}, F.~A.
  2002, \apj, 570, 171

\bibitem[{{Freitag} {et~al.}(2005){Freitag}, {G{\" u}rkan}, \&
  {Rasio}}]{2005astro.ph..3130F}
{Freitag}, M., {G{\" u}rkan}, M.~A., \& {Rasio}, F.~A. 2005, \mnras , submitted
  (astro-ph/0503130)

\bibitem[{{G{\" u}ltekin} {et~al.}(2004){G{\" u}ltekin}, {Miller}, \&
  {Hamilton}}]{2004ApJ...616..221G}
{G{\" u}ltekin}, K., {Miller}, M.~C., \& {Hamilton}, D.~P. 2004, \apj, 616, 221

\bibitem[{{G{\" u}rkan} {et~al.}(2004){G{\" u}rkan}, {Freitag}, \&
  {Rasio}}]{2004ApJ...604..632G}
{G{\" u}rkan}, M.~A., {Freitag}, M., \& {Rasio}, F.~A. 2004, \apj, 604, 632

\bibitem[{{Gebhardt} {et~al.}(2000){Gebhardt}, {Bender}, {Bower}, {Dressler},
  {Faber}, {Filippenko}, {Green}, {Grillmair}, {Ho}, {Kormendy}, {Lauer},
  {Magorrian}, {Pinkney}, {Richstone}, \& {Tremaine}}]{2000ApJ...539L..13G}
{Gebhardt}, K., {Bender}, R., {Bower}, G., {Dressler}, A., {Faber}, S.~M.,
  {Filippenko}, A.~V., {Green}, R., {Grillmair}, C., {Ho}, L.~C., {Kormendy},
  J., {Lauer}, T.~R., {Magorrian}, J., {Pinkney}, J., {Richstone}, D., \&
  {Tremaine}, S. 2000, \apjl, 539, L13

\bibitem[{{Gebhardt} {et~al.}(2002){Gebhardt}, {Rich}, \&
  {Ho}}]{2002ApJ...578L..41G}
{Gebhardt}, K., {Rich}, R.~M., \& {Ho}, L.~C. 2002, \apjl, 578, L41

\bibitem[{{Gebhardt} {et~al.}(2005){Gebhardt}, {Rich}, \& {Ho}}]{grh05}
---. 2005, \apj, submitted

\bibitem[{{Gerssen} {et~al.}(2002){Gerssen}, {van der Marel}, {Gebhardt},
  {Guhathakurta}, {Peterson}, \& {Pryor}}]{2002AJ....124.3270G}
{Gerssen}, J., {van der Marel}, R.~P., {Gebhardt}, K., {Guhathakurta}, P.,
  {Peterson}, R.~C., \& {Pryor}, C. 2002, \aj, 124, 3270

\bibitem[{{Gerssen} {et~al.}(2003){Gerssen}, {van der Marel}, {Gebhardt},
  {Guhathakurta}, {Peterson}, \& {Pryor}}]{2003AJ....125..376G}
---. 2003, \aj, 125, 376

\bibitem[{{Grandcl{\' e}ment} {et~al.}(2002){Grandcl{\' e}ment}, {Gourgoulhon},
  \& {Bonazzola}}]{2002PhRvD..65d4021G}
{Grandcl{\' e}ment}, P., {Gourgoulhon}, E., \& {Bonazzola}, S. 2002, \prd, 65,
  044021

\bibitem[{{Harris}(1996)}]{1996AJ....112.1487H}
{Harris}, W.~E. 1996, \aj, 112, 1487

\bibitem[{{Heggie}(1975)}]{1975MNRAS.173..729H}
{Heggie}, D.~C. 1975, \mnras, 173, 729

\bibitem[{{Hills}(1990)}]{1990AJ.....99..979H}
{Hills}, J.~G. 1990, \aj, 99, 979

\bibitem[{{Ivanova} {et~al.}(2005){Ivanova}, {Belczynski}, {Fregeau}, \&
  {Rasio}}]{2005MNRAS.358..572I}
{Ivanova}, N., {Belczynski}, K., {Fregeau}, J.~M., \& {Rasio}, F.~A. 2005,
  \mnras, 358, 572

\bibitem[{{Kalogera} {et~al.}(2004){Kalogera}, {King}, \&
  {Rasio}}]{2004ApJ...601L.171K}
{Kalogera}, V., {King}, A.~R., \& {Rasio}, F.~A. 2004, \apjl, 601, L171

\bibitem[{{Kidder} {et~al.}(1993){Kidder}, {Will}, \&
  {Wiseman}}]{1993PhRvD..47.3281K}
{Kidder}, L.~E., {Will}, C.~M., \& {Wiseman}, A.~G. 1993, \prd, 47, 3281

\bibitem[{{Kroupa} \& {Weidner}(2003)}]{2003ApJ...598.1076K}
{Kroupa}, P., \& {Weidner}, C. 2003, \apj, 598, 1076

\bibitem[{{Kulkarni} {et~al.}(1993){Kulkarni}, {Hut}, \&
  {McMillan}}]{1993Natur.364..421K}
{Kulkarni}, S.~R., {Hut}, P., \& {McMillan}, S. 1993, \nat, 364, 421

\bibitem[{{Lee}(1995)}]{1995MNRAS.272..605L}
{Lee}, H.~M. 1995, \mnras, 272, 605

\bibitem[{{Lee}(2001)}]{2001CQGra..18.3977L}
---. 2001, Classical and Quantum Gravity, 18, 3977

\bibitem[{{Lee}(1993)}]{1993ApJ...418..147L}
{Lee}, M.~H. 1993, \apj, 418, 147

\bibitem[{{Merritt} {et~al.}(2004){Merritt}, {Piatek}, {Portegies Zwart}, \&
  {Hemsendorf}}]{2004ApJ...608L..25M}
{Merritt}, D., {Piatek}, S., {Portegies Zwart}, S., \& {Hemsendorf}, M. 2004,
  \apjl, 608, L25

\bibitem[{{Miller} \& {Colbert}(2004)}]{2004IJMPD..13....1M}
{Miller}, M.~C., \& {Colbert}, E.~J.~M. 2004, International Journal of Modern
  Physics D, 13, 1

\bibitem[{{Miller} \& {Hamilton}(2002{\natexlab{a}})}]{2002ApJ...576..894M}
{Miller}, M.~C., \& {Hamilton}, D.~P. 2002{\natexlab{a}}, \apj, 576, 894

\bibitem[{{Miller} \& {Hamilton}(2002{\natexlab{b}})}]{2002MNRAS.330..232M}
---. 2002{\natexlab{b}}, \mnras, 330, 232

\bibitem[{{Murray} \& {Lin}(1996)}]{1996ApJ...467..728M}
{Murray}, S.~D., \& {Lin}, D.~N.~C. 1996, \apj, 467, 728

\bibitem[{{Peters}(1964)}]{1964PhRv..136.1224P}
{Peters}, P.~C. 1964, Physical Review, 136, 1224

\bibitem[{{Portegies Zwart} {et~al.}(2004){Portegies Zwart}, {Baumgardt},
  {Hut}, {Makino}, \& {McMillan}}]{2004Natur.428..724P}
{Portegies Zwart}, S.~F., {Baumgardt}, H., {Hut}, P., {Makino}, J., \&
  {McMillan}, S.~L.~W. 2004, \nat, 428, 724

\bibitem[{{Portegies Zwart} \& {McMillan}(2000)}]{2000ApJ...528L..17P}
{Portegies Zwart}, S.~F., \& {McMillan}, S.~L.~W. 2000, \apjl, 528, L17

\bibitem[{{Portegies Zwart} \& {McMillan}(2002)}]{2002ApJ...576..899P}
---. 2002, \apj, 576, 899

\bibitem[{{Quinlan} \& {Shapiro}(1989)}]{1989ApJ...343..725Q}
{Quinlan}, G.~D., \& {Shapiro}, S.~L. 1989, \apj, 343, 725

\bibitem[{{Schaller} {et~al.}(1992){Schaller}, {Schaerer}, {Meynet}, \&
  {Maeder}}]{1992A&AS...96..269S}
{Schaller}, G., {Schaerer}, D., {Meynet}, G., \& {Maeder}, A. 1992, \aaps, 96,
  269

\bibitem[{{Sigurdsson} \& {Hernquist}(1993)}]{1993Natur.364..423S}
{Sigurdsson}, S., \& {Hernquist}, L. 1993, \nat, 364, 423

\bibitem[{{Sigurdsson} \& {Phinney}(1993)}]{1993ApJ...415..631S}
{Sigurdsson}, S., \& {Phinney}, E.~S. 1993, \apj, 415, 631

\bibitem[{{Spergel} {et~al.}(2003){Spergel}, {Verde}, {Peiris}, {Komatsu},
  {Nolta}, {Bennett}, {Halpern}, {Hinshaw}, {Jarosik}, {Kogut}, {Limon},
  {Meyer}, {Page}, {Tucker}, {Weiland}, {Wollack}, \&
  {Wright}}]{2003ApJS..148..175S}
{Spergel}, D.~N., {Verde}, L., {Peiris}, H.~V., {Komatsu}, E., {Nolta}, M.~R.,
  {Bennett}, C.~L., {Halpern}, M., {Hinshaw}, G., {Jarosik}, N., {Kogut}, A.,
  {Limon}, M., {Meyer}, S.~S., {Page}, L., {Tucker}, G.~S., {Weiland}, J.~L.,
  {Wollack}, E., \& {Wright}, E.~L. 2003, \apjs, 148, 175

\bibitem[{{Spitzer}(1969)}]{1969ApJ...158L.139S}
{Spitzer}, L.~J. 1969, \apjl, 158, L139

\bibitem[{{Tutukov} \& {Yungelson}(1993)}]{1993MNRAS.260..675T}
{Tutukov}, A.~V., \& {Yungelson}, L.~R. 1993, \mnras, 260, 675

\bibitem[{{van der Marel}(2004)}]{2004cbhg.symp...37V}
{van der Marel}, R.~P. 2004, in Coevolution of Black Holes and Galaxies, 37

\bibitem[{{Watters} {et~al.}(2000){Watters}, {Joshi}, \&
  {Rasio}}]{2000ApJ...539..331W}
{Watters}, W.~A., {Joshi}, K.~J., \& {Rasio}, F.~A. 2000, \apj, 539, 331

\bibitem[{{Wen}(2003)}]{2003ApJ...598..419W}
{Wen}, L. 2003, \apj, 598, 419

\bibitem[{{White} \& {van Paradijs}(1996)}]{1996ApJ...473L..25W}
{White}, N.~E., \& {van Paradijs}, J. 1996, \apjl, 473, L25

\bibitem[{{Willems} {et~al.}(2005){Willems}, {Henninger}, {Levin}, {Ivanova},
  {Kalogera}, {McGhee}, {Timmes}, \& {Fryer}}]{2005ApJ...625..324W}
{Willems}, B., {Henninger}, M., {Levin}, T., {Ivanova}, N., {Kalogera}, V.,
  {McGhee}, K., {Timmes}, F.~X., \& {Fryer}, C.~L. 2005, \apj, 625, 324

\bibitem[{{Woitas} {et~al.}(2001){Woitas}, {Leinert}, \& {K{\"
  o}hler}}]{2001A&A...376..982W}
{Woitas}, J., {Leinert}, C., \& {K{\" o}hler}, R. 2001, \aap, 376, 982

\end{thebibliography}
\end{document}